\documentclass[a4,12pt,onecolumn]{article}
\usepackage[dvips]{graphicx}
\usepackage{amsmath,amssymb}
\newcommand{\Mat}[1]{{{\boldsymbol{#1}}}}

\def\be{\begin{equation}}
\def\ee{\end{equation}}
\def\dd{\mathrm{d}}
\date{}
\begin{document}
\title{\bf Ether theory of gravitation:\\ why and how?}

\author{Mayeul Arminjon}
\maketitle
\begin{center} { \small Laboratoire ``Sols, Solides, Structures, Risques'' \\
(CNRS \& Universit\'es de Grenoble), 
\\ BP 53, F-38041 Grenoble cedex 9, France.}

\vspace{0.5cm}

\begin{abstract}
Gravitation might make a preferred frame appear, and with it a clear space/time separation---the latter being, a priori, needed by quantum mechanics (QM) in curved space-time. Several models of gravitation with an ether are discussed: they assume metrical effects in an  heterogeneous ether and/or a Lorentz-symmetry breaking. One scalar model, starting from a semi-heuristic view of gravity as a pressure force, is detailed. It has been developed to a complete theory including continuum dynamics, cosmology, and links with electromagnetism and QM. To test the theory, an asymptotic scheme of post-Newtonian approximation has been built. That version of the theory which is discussed here predicts an internal-structure effect, even at the point-particle limit. The same might happen also in general relativity (GR) in some gauges, if one would use a similar scheme. Adjusting the equations of planetary motion on an ephemeris leaves a residual difference with it; one should adjust the equations using primary observations. The same effects on light rays are predicted as with GR, and a similar energy loss applies to binary pulsars.

\end{abstract}

\end{center}
\maketitle
\newpage
\tableofcontents
\vspace{2cm}
\section{Introduction}

The idea behind any attempt to build an ether theory is just that empty space ought not be really empty. We have two good reasons to think so: first, electromagnetic signals behave undoubtedly as waves; since they propagate even through intergalactic space, there must be some thing there (everywhere), in which they do wave. Second, quantum theory predicts that vacuum has physical effects, such as the Casimir effect, which is now experimentally confirmed \cite{Casimir}. However, we are taught at the University that the concept of the ether as the light-wave carrier, which had been dominating all Nineteenth Century physics, has been ruined by experiments such as that of Michelson and Morley, which have led to the development of special relativity (SR)---the latter abandoning, not only the apparently obvious notion that waves need a carrier, but even our basic intuition of simultaneity. That is, we are taught that one simply should forget common sense in view of the experimental evidence. It is a big surprise, therefore, to learn about an alternative version/interpretation of SR, according to which the Lorentz transform, and all the ``relativistic" effects involved in it, result actually from the ``absolute" Lorentz contraction experienced by any object that moves through a fundamental inertial frame or ether. This version of SR was initiated by the works of Lorentz \cite{Lorentz} and Poincar\'e \cite{Poincare1905, Poincare1906}. The latter work involves in fact the mathematical formalism and the basic physical content of SR. The Lorentz-Poincar\'e version of SR has been since investigated by Ives \cite{Ives}, Builder \cite{Builder}, J\'anossy \cite{Janossy}, Prokhovnik \cite{Prokhovnik67, Prokhovnik85}, Pierseaux \cite{Pierseaux}, and Brandes \cite{Brandes01}, among others. In particular, Prokhovnik \cite{Prokhovnik67, Prokhovnik85}, Pierseaux \cite{Pierseaux}, and Brandes \cite{Brandes01} proved in detail that this version is physically equivalent to the textbook version initiated by Einstein \cite{Einstein1905}. A brief vindication of the Lorentz-Poincar\'e version of SR has been recently presented by Morris \cite{Morris}.\\

The essential difference between the Lorentz-Poincar\'e version and the Einstein version is that the former version sees the key ``relativistic" effects (space contraction and time dilation) as real physical effects of the motion with respect to the ``ether": a moving ruler is just ``really" smaller than a fixed one, a moving clock just has a ``really" greater period than a fixed one. Here, ``really" means: with the simultaneity defined in the preferred inertial frame or ether E, by using the standard synchronization procedure with light signals (first proposed by Poincar\'e and rediscovered by Einstein). The reciprocity of these metrical effects (when interchanging the respective roles of two inertial frames) is then seen as an illusion due to the necessity of still using the Poincar\'e-Einstein synchronization convention in a moving frame: As long as we do not know the velocity $\mathbf{V}$ of our inertial frame F with respect to E, we are obliged to admit that, also in F, light has the same velocity in both directions $\mathbf{n}$ and $\mathbf{-n}$, for whatever $\mathbf{n}$ (that velocity must then be the constant $c$). What experiments like Michelson-Morley's ask for, and what the rod contraction and clock slowing say, is merely that the velocity of light {\it on a to-and-fro path} is the constant $c$. Thus, according to the Lorentz-Poincar\'e interpretation of SR, the metrical effects of uniform motion are seen as absolute effects due to the motion with respect to the preferred inertial frame, and the relativity of simultaneity is just an artefact. That artefact leads to the reciprocity of the ``relativistic" effects, which reciprocity is thus itself an artefact. Therefore, the Lorentz-Poincar\'e version of SR has the advantage over the standard (Einstein) version for the understanding of the relativistic effects and for the resolution of the allegated ``paradoxes" of SR (e.g. the clocks paradox and the Ehrenfest paradox, among many others); these advantages were demonstrated in detail by Prokhovnik \cite{Prokhovnik67}, and, in an even more convincing way, by Brandes \cite{Brandes01}.\\

However, if all of physics is really Lorentz-invariant, we shall never be able to measure the ``absolute" velocity $\mathbf{V}$, so that the difference between the ``true" simultaneity (that defined in the preferred inertial frame E) and the ``artificial" ones (those defined in the other inertial frames F) will remain at the metaphysical level. If this is the case in reality, then one may indeed qualify the concept of ether as ``superfluous," as Einstein \cite{Einstein1905} put it. His own starting point, from the relativity principle, is then to be preferred. Moreover, the full equivalence between all inertial frames, obtained in that case, leads to give a physical meaning to the concept of space-time, which was Minkowski's assumption \cite{Minkowski1907}. (In his ``Rendiconti" paper, Poincar\'e \cite{Poincare1906} had already introduced the concept of the space-time, as a 4-dimensional space with coordinates $x, y, z, ict$ (and $c=1$ in his paper); but this looked more like a (very useful and clever) mathematical tool in his paper.) This means that past, present and future are relative notions, and that time travels are at least theoretically possible.
\footnote
{~In general relativity, closed time-like curves (involving the possibility of a time travel) do occur. According to Bonnor \cite{Bonnor02}, ``They can no longer be dismissed as curiosities occurring in non-physical solutions. We now know that there are simple physical situations, such as that of the charge and the magnet, within the terms of reference of general relativity, of which the theory as currently understood does not give a satisfactory account." 
}
 It also enforces to consider this mixture of space and time as the true arena for physical theories. In particular, it may lead (though not in a compelling way) to Einstein's general relativity (GR), according to which gravitation is the curvature of space-time, and in which free test particles follow the geodesics of a Lorentzian metric on space-time. 
In my opinion, this is no physical explanation of gravity. Moreover, GR's invariance under arbitrary coordinate changes is antinomic to quantum theory, in which, in particular, the choice of the {\it time} coordinate cannot be arbitrary \cite{B15}. This leads to the longstanding difficulties with quantum gravity, which the mainstream researchers want to overcome by going to string- and M-theories with their many-dimensional manifolds---which represents a jump to still orders-of-magnitude more complex theories, as compared with GR which is already very complex. One should be allowed to explore simpler possibilities.\\

There is indeed another possibility, which seems to have been little explored before: that not all of physics is subjected to the relativity principle, and that the force which contradicts relativity is precisely gravitation. (Of course, the conflict with relativity has to be small in usual conditions.) This possibility exists only in the Lorentz-Poincar\'e version of SR, that sees SR as a consequence of the Lorentz contraction: it obviously does not exist if the reason for SR is the universal validity of the relativity principle, as in Einstein's version. Why should gravitation be the range of a such violation? Simply because SR does not include gravitation. In GR, the presence of a gravitational field means that the space-time manifold, say V (which is a Lorentzian manifold, i.e., it is endowed with a Lorentzian metric $\Mat{\gamma}$), is not flat any more, so that the Lorentz group acts only ``locally" on V, instead of truly acting on V as it is the case in SR. 
\footnote
{~When one says that the Lorentz group G = O$(1,3)$ acts ``locally" (or rather infinitesimally) on V, one actually just means that this group operates on the tangent space $\mathrm{TV}_X$ to V at any point $X\in \mathrm{V}$ (and indeed $\mathrm{TV}_X$ is invariant under the action of any element of G). This does not restrict in any way the generality of the Lorentzian manifold $(\mathrm{V},\Mat{\gamma})$. Whereas, in this jargon, the relativity principle of SR may be expressed by saying that V is identical to (or may be identified with) any of its tangent spaces (which again are invariant under the action of any element of G), that is indeed a very strong restriction.
}
Thus, in GR, the relativity principle of SR simply does not hold true. This has been noted a long time ago by Fock \cite{Fock59}. The ``general relativity" is the fact that, under {\it general} coordinate changes, GR is ``{\it manifestly}" covariant (i.e. manifestly form-invariant), thus without introducing any extraneous variables such as a velocity. It means physically that there is no preferred reference frame in GR, and even that there is no preferred {\it class} of reference frames (whereas the latter is the case in SR), and this is of course very important. But, if one takes the Lorentz-Poincar\'e interpretation of SR seriously, one is allowed to ask whether the extension of that theory to include gravitation ought no to be done in the opposite way, namely by assuming that there {\it is} in fact a preferred reference frame, which just remains ``hidden" when gravity can be neglected.\\

The availability of a ``preferred foliation of space-time" (in the words of Butterfield \& Isham \cite{Butterfield}), i.e., of a preferred reference frame, would provide a perspective of solution to the problems at the interface between quantum theory and gravitation. Quantum theory, being based on a generalization of hamiltonian mechanics, should a priori need a preferred time coordinate. In the case of a {\it flat} space-time, the ``inertial time" is preferred, that is simply the Poincar\'e-Einstein synchronized time. It is not a too serious problem that every inertial frame has its specific inertial time, because the canonical quantization rules lead to Lorentz-invariant equations---or, at least, can lead to Lorentz-invariant equations, this condition becoming then an a priori constraint on the final form of the quantum equations. However, in the case of a space-time assumed to be curved by gravitation, there is no way to distinguish one time coordinate among the infinitely many possible ones, except for some particular situations (notably the case of a static gravitational field). Anyway, to distinguish one particular time coordinate for the purpose of unambiguously formulating quantum equations and quantum principles would contradict the very principle of ``general relativity" and would lead to equations having a smaller covariance group, an inacceptable feature according to GR. In contrast, this process would be completely admissible if one would start from a preferred-frame theory of gravitation. The main difficulty is then, of course, to build a such theory that would agree with observations. 

\section{A few models of gravitation consistent with the Lorentz-Poincar\'e ether theory}

\subsection{The models of Podlaha, Sj\"odin, and Broekaert} \label{podlaha}

Podlaha \& Sj\"odin \cite{PodlahaSjodin} considered an ether characterized by its density and velocity fields, and formulated a ``principle of equivalence of second order," according to which ``the effects caused by a real static gravitational field, i.e. by a static inhomogeneous aether field, are observationally equivalent to the effects `caused' by the apparent universal velocity field"---the latter arising from the motion, relative to the given observer, of all bodies bound to the ether, which motion the observer may infer from the fact that ``he will observe some phenomena due to this motion, as e.g. the contraction of bodies and the slowing down of the rates of clocks." The formulae deduced from this principle ``indicate the slowing down of the rates of clocks and contraction of bodies at rest in the aether." They led Podlaha \& Sj\"odin \cite{PodlahaSjodin}, in the static spherically symmetric (SSS) case, to an approximate relation between the ether density and the Newtonian potential, making the former increase in the direction of the attracting center. By a priori assuming the latter relation, Sj\"odin \cite{Sjodin1982} had been able to interpret the gravitational light deflection in the gravitational field of a spherical body as due to the propagation of light in a medium with space-variable refraction index, that index being assumed to be just the ether density. Later, Sj\"odin \cite{Sjodin1990} introduced a Lagrangian for a massive test particle in an SSS gravitational field, involving in general two functions $\Phi$ and $\Omega$ characterizing respectively the ``change of dimensions [of] material, measuring rods" and the ``change of frequency [of] real, material clocks" in the static gravitational field, both functions being primarily ones of the ``refractive index for light in the physical vacuum (`aether')". According to his calculation of the motion of particles derived from this Lagrangian, this motion would be the same as that deduced in GR from geodesic motion in the Schwarzschild metric, if a specific form $\Phi\neq\Omega$ was assumed, which form turns out to be different from that expected from Ref. \cite{PodlahaSjodin} (this expectation includes indeed the relation $\Phi=\Omega$).\\

Recently, Broekaert \cite{Broekaert} has proposed a Hamiltonian for a static gravitational field. Although he does not refer to the concept of ether, his Hamiltonian involves a ``static gravitational affectation function" $\Phi$, defined in the general (not necessarily spherical) static case, and which indeed plays much the same role as Sj\"odin's function $\Phi$, with furthermore $\Phi=\Omega$, and ``the corresponding Lagrangian [is] the Lagrangian used by Sj\"odin" in Ref.~\cite{Sjodin1990} here. However, based on a ``gravitationally modified Lorentz transformation" in which ``the invariance of the locally observed velocity of light is secured by adequately balancing the variability of speed of light and gravitational affectation of space and time measurement," and using a ``Newtonian fit" by which an observer at rest relative to the background field $\Phi$ ``can identify the change of static energy as a shift in Newtonian potential energy," he finds (again in the general static case) a nonlinear equation for the function $\Phi$. This equation is solved by a quadrature in the general static case, and explicitly for an SSS field. In the SSS situation, Broekaert \cite{Broekaert} finds then that the predictions for the deflection of light rays (particles with zero rest-mass), the perihelion precession for a massive test particle in a bounded orbit, the radar echo delay, and the gravitational redshift, are identical (to first post-Newtonian order) to the predictions deduced in GR from Schwarzschild's metric.\\

Thus, thanks to this chain of work \cite{PodlahaSjodin,Sjodin1982,Sjodin1990,Broekaert}, an alternative dynamics of test particles has been defined in a static gravitational field, starting from the consideration of a scalar field (the ``aether density") which, consistently with Lorentz-Poincar\'e SR, is assumed to ``affect" our physical space and time standards. It is difficult to guess whether this could be the beginning of a complete theory of gravity. Even for solar system tests, it is not enough to consider the SSS situation (this has been emphasized by Will \cite{Will}), essentially because, in celestial mechanics, one must account for the fact that ``the solar system is {\it not} static and isotropic"---as emphasized by Weinberg \cite{Weinberg}. In fact, in celestial mechanics, the main corrections to the Newtonian SSS case come from the Newtonian perturbations due to the planets (which are Newtonian corrections {\it to} the SSS case), not from the post-Newtonian effects {\it within} the SSS situation. But it is simply incorrect to brutally add Newtonian perturbations to the post-Newtonian solution of the SSS case: one has first to formulate the complete problem in the framework of a new theory and {\it then} introduce {\it numerically justified} approximations, instead of {\it ad hoc} ones; this was noted by Feyerabend \cite{Feyerabend}, who qualified the first (brutal) procedure a ``methodological nightmare". As it will appear in this paper, it is quite a work to go even from a theory already formulated in the general case to numerical predictions deduced from consistently derived post-Newtonian equations for a weakly gravitating system of {\it extended} bodies. (In retarded and/or nonlinear theories, it is very important to account for the finite extension of the bodies, and at the same time this is dangerous for such theories.
\footnote{
~Indeed the fact that the acceleration of the mass center of a body depends only on the other bodies in Newtonian gravity, depends crucially on the actio-reactio principle. That principle does not hold true for retarded theories \cite{Poincare1900}, and it is meaningless for nonlinear ones since, for a nonlinear theory, one cannot even define what could be the action ``due to a given part of the system". As a consequence, one may expect that each of the mass centers will, in general, be subjected to a self-force, which is the dangerous point.
} \label{actioreactio}
But, before doing this, it is of course necessary to {\it define} the dynamics for an extended body.) The foregoing remarks mean that there is currently no celestial mechanics associated with Broekaert's model \cite{Broekaert}. In particular, the result found for a test particle in an SSS field, while encouraging, does not allow to state that some extension of this model might actually account for Mercury's advance in perihelion in a realistic celestial-mechanical model. The situation is better for gravitational effects on light rays in Broekaert's model, because for light rays it seems justified to consider a unique spherical body---but this body can have a motion through the preferred frame if the considered theory is a preferred-frame theory, in which case it is necessary to study the effects of that motion \cite{A19}. However, one cannot say in advance whether or not a possible future extension of Broekaert's model to a general situation would lead to a preferred-frame theory.
\footnote{
After the first version of the present work was completed, Broekaert has extended his model by introducing a ``sweep velocity field" \cite{Broekaert2004} that mimics the vector potential which occurs in the equations of PN GR in the harmonic gauge. This makes it possible to consider more general situations, such as that of a rotating source. For the latter, the Lense-Thirring effect is reproduced, according to Broekaert \cite{Broekaert2004}. The dynamics of the model remains restricted to test particles.
}

\subsection{Winterberg's theory} \label{winterberg}

Another model of gravitation in the line of Lorentz-Poincar\'e relativity is Winterberg's vector theory of gravity with substratum \cite{Winterberg}. In this theory with a flat space-time, the gravitational field is defined, in analogy with electromagnetism, by a scalar potential $\Phi$ and a 3-vector potential $\mathbf{A}$, and its source is ``the special-relativistic four-current vector of matter"---the rest mass density $\rho_0$ being, however, that ``of the sources as they would appear in the absence of a gravitational field," an expression which should be precised in order to make it clear which field is the actual source of the gravitational field in the actual space-time including the bodies and the gravitational field. Furthermore, ``the gravitational field is assumed to set into motion the substratum" (ether), indeed the substratum obeys the Newtonian equation of motion for a continuum subjected to a kind of Lorentz force deduced from the gravitational potentials $\Phi$ and $\mathbf{A}$, though involving a coefficient $\alpha$. The latter is found to have to be equal to 4 in order that the Newtonian motion of the substratum be equivalent, to first order, to a geodesic motion in the space-time endowed with a curved Lorentzian metric obtained as a small perturbation, involving the potentials $\Phi$ and $\mathbf{A}$, of the flat metric. Since $\Phi$ and $\mathbf{A}$ obey usual wave equations whose source is the four-current of matter (Eqs. (3.1) in Ref.~\cite{Winterberg}), this coefficient $\alpha =4$ means, as noted by Winterberg, that the gauge invariance of $\Phi$ and $\mathbf{A}$ is lost. He insists that ``the solution of the gravitational field equations, however, must still be invariant under a Lorentz transformation" defined by a constant velocity $\mathbf{v}_0$ (a 3-vector), and he states that this is obtained by setting $\mathbf{A}'=\mathbf{A}-\small{\frac{1}{4}}c\mathbf{v}_0$ and $\Phi'=\Phi-\small{\frac{1}{2}}\mathbf{v}_0^2$. However, this is not the usual way to Lorentz-transform a 4-vector (which is here $\mathbf{\mathsf{A}}\equiv (\Phi,\mathbf{A})$), moreover one has to define also the transformation of the source term, which is not done in Ref.~\cite{Winterberg}. There is in fact an obvious way to transform wave equations for a 4-potential $\mathbf{\mathsf{A}}$ having as its source the {\it usual} four-current of matter, $\mathbf{\mathsf{t}} \equiv \rho^*\mathbf{\mathsf{U}}$ (where $\rho^*$ is the {\it proper} rest-mass density, which is an invariant scalar, and $\mathbf{\mathsf{U}}$ is the 4-velocity), and this is the usual Lorentz transform of these two 4-vectors (i.e., just the same transformation as that which applies to the space-time coordinates). Obviously also, this is not the way chosen by Winterberg \cite{Winterberg}. Hence, it is not clear in which sense it is claimed that equations (3.1) of Ref.~\cite{Winterberg} are ``invariant under a Lorentz transformation," which is the corresponding transformation of the source term---and, as mentioned above, how is this source term precisely defined. 
\footnote{
~If the equations for the 4-potential $\mathbf{\mathsf{A}}$ are taken to be $\square A^\mu =t^\mu $, then of course they are covariant under Lorentz transforms, even they are generally-covariant. (Contrary to Winterberg, I use Latin letters for spatial indices and Greek letters for space-time indices.) The reason why this is (definitely) not Winterberg's \cite{Winterberg} choice may be related with the fact that he postulates a {\it Newtonian} equation of motion for the substratum, which a priori does not fit well with (usual) Lorentz invariance. It may be that once the density is precisely defined in Eqs.~(3.1) of Ref.~\cite{Winterberg}, a preferred frame will appear.
}
\\

An essential feature in Winterberg's theory \cite{Winterberg} is his version of the principle of equivalence, according to which version ``we can determine the forces acting on a body if placed in the flowing substratum set into motion through a gravitational field, by comparing them with the inertial forces which result if the body is placed in an accelerated frame of reference, relative to which the substratum is in motion." This assumption leads him to state that ``the effect of the substratum motion on rods and clocks in an accelerated frame of reference, as in the ether interpretation of special relativity, is uniquely determined by the motion of the substratum. The substratum velocity in the accelerated frame of reference here too causes the rod contraction and time dilation effects." This way of reasoning leads him finally to state his Eqs. (4.16), (4.22) and (4.24), determining a curved Lorentzian metric  $\Mat{\gamma}$, in terms of the flat metric and the velocity field of the substratum (but the three components $\gamma_{0i}$ remain implicit in the general case). The latter field has to be obtained, I recall, as a solution of the Newtonian equation of motion for a continuum subjected to the Lorentz-type force deduced from the gravitational potentials $\Phi$ and $\mathbf{A}$. The latter ones, conversely, are determined from the current of matter through the equations discussed in the foregoing paragraph. That current describes the motion of the bodies, which itself should be determined from general-relativistic-type equations of motion in the metric $\Mat{\gamma}$.
\footnote{
~In fact, Winterberg defines explicitly only the equation of motion for a test particle (section 5 in Ref.~\cite{Winterberg}), and this he defines indeed as the variational equation for the geodesic lines of the metric $ds^2$ (his Eq. (3.10)), ``in exactly the same way as it is done in Einstein's theory." Therefore, one may guess that the motion of extended bodies should also be defined from the usual equation of GR, $T^{\mu \nu }_{\quad;\nu }=0$ with $\mathbf{T}$ the energy-momentum tensor. Actually, the latter equation is equivalent to geodesic motion for a dust. The universality of gravity should then indeed enforce to keep the same equation for a general matter behaviour.
}
 
Thus, Winterberg's theory is expressed by a complex system of coupled nonlinear partial differential equations, which might be described as a kind of magnetohydrodynamics with two mutually interpenetrating fluids: the substratum, which is indeed subjected to an MHD, and the material fluid, which is a relativistic fluid in curved space-time---in case matter is schematized as a fluid. 
\footnote{
~If matter is schematized as a {\it barotropic} perfect fluid (thus eliminating the pressure), I see in Winterberg's theory 15 unknown fields, instead of 14 in GR and 5 for the scalar theory~\cite{A20}: $\Phi$, $A^i$, $\rho_0$, $V^i$, $v^i$, $g_{00}$, $g^i$. And I see 15 independent equations: 4 (not 5) in (3.1)-(3.2), 3 in (3.6), 1 in (4.22), 3 in (4.24), and (see Note 5) the four ones in $T^{\mu \nu }_{\quad;\nu }=0$. 
}
In the SSS case, Winterberg finds that the curved metric outside the spherical body is the Schwarzschild metric (whereas the metric inside an incompressible fluid sphere differs from Schwarzschild's interior solution). For the reasons given in the last part of subsection \ref{podlaha}, this makes it {\it plausible} that his theory may explain gravitational effects on light rays, but it does not allow to state that Mercury's advance in perihelion is explained by this theory: one would have to develop the post-Newtonian (PN) celestial mechanics of extended bodies in that theory, and this would certainly be a difficult task. If one attacks it, he will have first to check whether there is a correct Newtonian limit in this theory (this is the zero-order PN approximation \cite{Rendall92, B18}). A theory that would not have a correct Newtonian limit for, say, a perfect-fluid system, and this in a general situation, would have no chance to be viable. In Ref.~\cite{Winterberg}, the Newtonian limit is touched on only for a test particle in the static case (Eqs. (5.3-5) there). I think that the Newtonian limit should be studied in detail in Winterberg's theory before checking further features of that theory~\cite{Winterberg}, regarding e.g. the avoidance of the singularities. 

\subsection{Gravitation in a solid-state substratum or in a Higgs condensate}

Dmitriyev \cite{Dmitriyev92} considers physical vacuum as a linear-elastic continuum, in an attempt to show that ``physical space, matter, and interactions can be interpreted as the phenomenology of a solid-state substratum [...] The transverse wave of [the] physical vacuum is a model of electromagnetic radiation (light), while the longitudinal wave is a model of a gravitational wave." He assumes that the velocities $c_r$ and $c_g$ of electromagnetic and gravitational signals verify $c_r\ll c_g$, and he adds: ``In terms of material constants, the condition $c_r\ll c_g$ means that [the] physical vacuum is an almost-incompressible medium, but one that is compliant to shear strain. Among model constructs, this property is exhibited by packing of identical solid spheres." Particles of matter are modeled as point defects in the linear solid medium, more precisely as ``dilatation centers": each such point gives rise to a displacement field whose divergence is a Dirac measure at this point. A distribution of such centers, [each of] ``which can move freely in the volume of [the] homogeneous solid continuum" (without any transport of material of the medium), is then ``consider[ed] as a kind of fluid, specifically a point dilatation fluid. [...] A Newtonian gravitational field is modeled by a constant convective flux of the point dilatation fluid, emitted by a blob of this fluid." A calculation of ``the interaction energy of two weak point sources of this flux" leads him to an expression of the form Const.$b_1 b_2/R$, where $R$ is the distance between the two sources and $b$ is the total power of an inclusion (a point source of the flux), which, ``in this case, [...] can be interpreted as the gravitational mass of a particle of matter." Thus, that expression may be interpreted as the Newtonian gravitational energy for a system of two point masses.\\

We can see that, in the framework of his substratum model, Dmitriyev \cite{Dmitriyev92} proposes an analogue model of gravitation, rather than a new phenomenological model of gravity. (This analogue model of gravity coexists with analogue models for the other interactions, in the framework of the ``vortex sponge" model \cite{Dmitriyev93,Dmitriyev99}---also advocated by Duffy \cite{Duffy2004} as a promising unifying concept for physics, in a hierarchical model of ether.) However, his model favours a nearly instantaneous propagation for gravity (as compared with light), which certainly is a definite qualitative prediction. This prediction seems very dangerous in view of the numerous observational confirmations of GR (a theory which states that $c_r = c_g$, in Dmitriyev's notation). It is interesting, however, to recall that Van Flandern \cite{Van Flandern} has studied a model of gravitation with propagation, based on a priori assuming an ``aberration formula" for the gravitation force and, therefore, leading to {\it first-order} departure from Newton's theory (i.e., like $v/c_g$ with $v$ the typical velocity)---as is also the case for the (different) force law proposed by Beckmann \cite{Beckmann}. If this model with first-order retardation effects \cite{Van Flandern} is taken to be a correct model, observations in the solar system and observations of binary pulsars would imply that ``the speed of gravity is no less than $2.10^{10}c$"... In that case, we would be facing a dramatic crisis, since it would mean that there is a deadly clash between gravitation and relativity. But the observations quoted by Van Flandern do not lead us to be confronted with that crisis, because several theories of gravitation accounting for SR, and stating that $c_g=c$, or at least $c_g \simeq c$, lead only to {\it second-order} ($v^2/c^2$, thus very small) departure from Newton's theory. This applies to GR \cite{Rendall92}, and it has been proved quite in detail~\cite{B18,A23} in the scalar theory that will be summarized in Section 3 here.\\

It is striking to note the connection of Dmitriyev's model \cite{Dmitriyev92}, interpreting gravitation as the propagation of a distribution of dilatational point defects in a solid continuum, with the work of Consoli \cite{Consoli_condensate}, although the latter is formulated in the language of quantum field theory. Consoli ``discuss[es] some phenomenological aspects of the ground state of spontaneously broken theories: the 'Higgs condensate'. The name itself [...] indicates that, in our view, this represents a kind of medium made up by the physical condensation process of some elementary quanta. If this were true, such a vacuum should support long-wavelength density fluctuations. In fact, the existence of density fluctuations in any known medium is a basic experimental fact depending on the coherent response of the elementary constituents to disturbances whose wavelength is much larger than their mean free path. [...] Some arguments suggest that, indeed, the vacuum of a 'pro forma' Lorentz-invariant quantum field theory may be such kind of medium. For instance, a fundamental phenomenon as the macroscopic occupation of the same quantum state (say $\mathbf{p = 0}$ in some frame) may represent the operative construction of a 'quantum aether' [...] This would be quite distinct from the aether of classical physics, considered a truly preferred reference frame and whose constituents were assumed to follow definite space-time trajectories. However, it would also be different from the empty space-time of special relativity, assumed at the base of axiomatic quantum field theory to deduce the exact Lorentz-covariance of the energy spectrum. In addition, one should take into account the approximate nature of locality in cutoff-dependent quantum field theories. In this picture, the elementary quanta are treated as 'hard spheres', as for the molecules of ordinary matter." (Compare my quote from Dmitriyev \cite{Dmitriyev92} above.) He finds that (in the relevant limit), ``long-wavelength density fluctuations would propagate instantaneously in the spontaneously broken vacuum" and he suggests ``the possible relevance of our picture in connection with the problem of gravity," whereby in fact gravity would propagate nearly instantaneously---as was more explicitly stated in Ref.~\cite{Consoli_grav_scal}.\\

In summary, in the solid-state analogue model of Dmitriyev~\cite{Dmitriyev92}, as well as in the considerations of Consoli~\cite{Consoli_condensate, Consoli_grav_scal} on a field-theoretical ``condensate," it is found that gravity should be related with the propagation of density fluctuations in some medium, which propagation should occur with a velocity $c_g$ much larger than light. This is not {\it forbidden} by the excellent agreement between observations in the solar system and GR (as operated using the standard PN approximation scheme \cite{Fock59,Chandra65}), because the PN scheme predicts that the finite propagation speed has no effect up to the order $1/c^2$, and in fact until $1/c^4$ included. However, it would remain to state precisely which modification of Newtonian gravity, compatible with $c_g\gg c$, might indeed predict the same observational agreement: unlike Consoli \cite{Consoli_grav_scal}, I believe that the ``Equivalence Principle" is not enough to ensure that ``the classical tests in weak gravitational fields are fulfilled as in general relativity." In my opinion, a theory has to be ultimately characterized by a definite set of equations, not by a such ``principle". Moreover, a propagation speed $c_g\gg c$ does contradict the accepted interpretation of binary-pulsar observations as showing an energy loss connected with the emission of gravitational waves at speed $c$ \cite{TaylorWeisberg}. If gravity would propagate with a speed $c_g\gg c$, these very accurate observations would have to find a totally different interpretation.

\subsection{Provisional conclusion}

A good number of attempts at an ``ether theory of gravity" have been presented in the last two decades. Several of them aimed explicitly at extending the Lorentz-Poincar\'e ``ether version" of SR to the situation with gravity \cite{PodlahaSjodin,Sjodin1982,Sjodin1990,Winterberg,Schmelzer}. However, also the works which do not start from the Lorentz-Poincar\'e version of SR \cite{Broekaert,Dmitriyev,Consoli_condensate,Consoli_grav_scal,JacobsonMattingly} are consistent with it much more than they can be with conventional SR.~\footnote{
~I apologize for quoting the theories of Schmelzer~\cite{Schmelzer} and of Jacobson \& Mattingly~\cite{JacobsonMattingly} without discussing them. Although the latter is truly an ``ether theory," it remains quite close to GR as compared with the other models quoted. Schmelzer's theory is also close to GR. In fact, in most respects, it is nearly identical to the RTG (relativistic theory of gravitation) of Logunov et al.~\cite{Logunov88,Logunov89}, in which the harmonic gauge condition often used in GR becomes a (generally-covariant) additional field equation and a massive graviton is assumed---although Schmelzer's theory involves an amazing rewriting of the harmonic condition as the Newtonian dynamical equations for a continuum without external force. I prefer to focus on strongly alternative theories.
}
Indeed they all introduce some form of breaking of the exact Lorentz symmetry, and this is hard to justify in the framework of Einstein's approach to SR: since, in this approach, the reason for SR is the universal validity of the relativity principle, there is no room there for a violation of this principle. In contrast, if one studies physically possible transformations of space and time coordinates without a priori imposing the relativity principle, one finds that the exact Lorentz symmetry is an ``unstable" point in the spectrum of all possible transformations: as demonstrated by Selleri \cite{Selleri91}, ``any violation of the conditions [for the exact Lorentz transform], however small, leads necessarily to an ether theory."

It appears from the foregoing review that these theories remain currently at an early stage in their development. E.g., even for those which are already complete, no PN celestial mechanics is available, in fact not even a complete Newtonian limit. (This is even true for Schmelzer's theory~\cite{Schmelzer}, insofar as it does not coincide with harmonic GR: one should investigate the effect of the additional terms in its Lagrangian, as compared with the GR Lagrangian. Even if one wishes that these terms have a negligible effect, it remains to chart the corresponding domain in the parameter space.) In the sequel of this paper, I will discuss in some detail my own attempt, which I have developed rather extensively. (However, few formulae will be given here.) This development was a long work, but I could not have done it, if my attempt had not been simple enough---in fact it is based on just one scalar field. 

\section{A scalar ether theory of gravitation}

\subsection{Gravitation as a pressure force} \label{grav=pression}

The construction of the theory begins with an interpretation of gravity, in the framework of {\it classical mechanics}, as ``Archimedes' thrust in the ether," which would take place at the scale of elementary particles~\cite{A8}. I learned since that Newtonian gravity (NG) had been interpreted in that way, and this in some detail, by Euler \cite{Euler}. In my work, this interpretation leads to a generalization of NG, already in the framework of classical mechanics, as soon as one endows the ether with a compressibility. My interpretation may be criticized, mainly because our present understanding of particle physics makes it quite doubtful that classical mechanics could apply at this scale. But who knows? Moreover, a theory is, after all, contained in its final equations, which may be considered in an axiomatic way---it is precluded that one may ``prove the equations" of a new physical theory. The following can thus be regarded as a heuristic approach to the equations. There is in fact another, phenomenological way, to obtain Eq.~(\ref{vecteur_g}) below, that gives the gravity acceleration: see Ref.~\cite{A16}, Subsect. 3.1. However, I do not disavow that beginning, and I am now going to summarize it with enough detail.\\

The first point is that, in order that it does not brake the motion of material bodies, the physical vacuum or ``micro-ether" must be some kind of a perfect fluid. 
\footnote{
~But, on the other hand, the ether should define an inertial frame, because postulating an ether plus an independent absolute space seems too much. My solution to this problem is that it is the {\it average motion} of the micro-ether that defines an inertial frame~\cite{A15,A28}.
}
A ``truly perfect" fluid is free from any thermal effect that is necessarily bound to dissipation, hence, as noted by Romani~\cite{Romani}, it must be perfectly continuous at any scale. It is then characterized by its pressure and its density, which are connected by the state equation, and by its velocity. It exerts only pressure forces. Therefore, if one attempts to introduce a perfectly fluid ether ``filling empty space," then any interaction forces ``at a distance," thus including gravity, have to be ultimately explained as pressure forces, and hence as contact actions. As far as gravitation is concerned, this is quite simple. I assume that elementary particles are extended objects. The resultant of the pressure forces exerted on a particle is Archimedes' thrust, that is proportional to the volume $\delta V$ occupied by a given particle. In order that this force be actually proportional to the {\it mass}~ $\delta m$ of the particle, it is hence necessary and sufficient that the average density inside a given particle, thus $\rho_p=\delta m/\delta V$, be the same for all particles---at least at a given (macroscopic) place and at a given time. However, since the gravitational attraction is a field, the density $\rho_p$ may also be a field, whose the space-time variability has to come from that of the pressure in the fluid, $p_e$. In fact, as is suggested by the observed transmutations of elementary particles into different ones, I assume that the particles themselves are made of that micro-ether: each of them should be some kind of organized flow in this imagined fluid---something like a vortex. (This is Romani's idea of a ``constitutive ether"~\cite{Romani}.) In that case, the density $\rho _p$ would be nothing else than the local density in the fluid, $\rho _e = \rho _e(p_e)$. Under these assumptions, the gravity acceleration is obtained as~\cite{A8}: 

\begin{equation} \label{vecteur_g}
{\mathbf{g}}= -\frac{\mathrm{grad}\: p_e}{\rho_e}. 
\end{equation}

Note that this relation implies that the ``ether pressure" $p_e$ and its density $\rho_e$ {\it decrease} in the direction of the gravitational attraction. Since the latter varies only on a macroscopic scale, the fields $p_e$ and $\rho_e$ in Eq.~(\ref{vecteur_g}) must actually be the {\it macroscopic} fields, obtained by space-averaging the microscopic fields. Hence, these microscopic fields should be connected with the other interactions, that vary over shorter distances. Also note that gravity is thus defined as a kind of {\it correction} (specifically that arising from a ``residual" pressure gradient at the macro-scale), which helps to understand why it is in fact so weak.\\

Further, I admit that {\it Newtonian gravity, because it propagates instantaneously, must correspond to the case of an incompressible fluid}. 
\footnote{
~This was already suggested by Romani~\cite{Romani}, although he assumed that gravity was due to an {\it increase} of ether density towards the attracting center, thus as Podlaha \& Sj\"odin \cite{PodlahaSjodin}, and he did not have anything like Eq.~(\ref{vecteur_g}).
}
By imposing that $\mathbf{g}$ given by Eq.~(\ref{vecteur_g}) should satisfy the Poisson equation when $\rho_e$ is uniform, one gets $\Delta p_e = 4 \pi G \rho \rho_e$, where $\Delta$ is the usual (flat) Laplacian and $\rho$ is the density of usual matter. ($G$ is the gravitation constant.) That equation still makes sense in the case that, due to the compressibility of the fluid, $\rho_e=\rho_e(p_e)$ is not uniform any more. However, if the ether is a compressible fluid, that equation can only apply in the static case. In the general case, indeed, we expect that there should be pressure waves, thus {\it gravitational waves}, which should propagate with the ``sound" velocity in the ether,
\be    \label{c_e}
c_e = {\left(\frac{\dd p_e}{\dd \rho_e}\right)}^{1/2}.
\ee
This is indeed what is found~\cite{A8} by adapting the classical reasoning that leads to the acoustic wave equation, if one admits, in particular, that the ether is conserved and that the average motion of the ether with respect to its mean rest frame is governed by Newton's second law. (That motion is the space-averaged motion, since such are the fields $p_e$ and $\rho_e$. As to the ``mean rest frame of the ether," noted E, and which is the preferred reference frame of the theory, its definition involves a time average, in addition to the space average~\cite{A28}.) In that way, one gets, in the general case, a wave equation involving the (variable) speed $c_e$ for the field $p_e$, the r.h.s. being still $4 \pi G \rho \rho_e$---which already makes the theory {\it nonlinear}. The SSS solution tends towards the Newtonian solution in the limit of weak fields, and, for a test particle in a bound orbit, it leads to an advance in perihelion which is proportional to that obtained using the Schwarzschild solution of GR~\cite{A8}. 

\subsection{Equivalence principle between metrical effects of motion and gravitation} \label{Metric}

Since the theory starts from classical mechanics, we have to adapt it in order to account for relativistic effects, and this is easier if we adopt the Lorentz-Poincar\'e version of SR, recalled in the Introduction. As is known, in GR, a fundamental role is played by Einstein's principle of equivalence between the effects of inertial and gravitational forces. When one studies Einstein's well-known examples of the lift and the rotating disk, one may get the impression that this principle is actually one that provides a relation between the metrical effects of uniform motion and those of gravitation. The essential conclusion which is drawn from these examples is, indeed, that there are metrical effects of a gravitational field, and that, for a weak field, these effects should be related to the local value of the Newtonian potential $U$, that plays the role played in SR by the squared velocity $v^2$. However, according to the Einstein version of SR, the metrical effects of a uniform motion are merely a kind of ``parallax in space-time," they depend on the reference frame. It can then not be envisaged to formally deduce the metrical effects of gravitation from those of uniform motion, because the former ones should have an absolute character: even in GR, the presence or absence of a gravitational field (i.e., of a non-zero Riemann-Christoffel tensor) is absolute. In GR, the essential role of the equivalence principle is to introduce the basic idea of GR, that a gravitational field can be characterized by a non-vanishing curvature (tensor) in a Lorentzian space-time manifold. According to Synge~\cite{Synge64}, this principle was inaccurate and had just had the role of a heuristic tool at the step of the construction of GR. For him, this principle was simply to be abandoned since we have the Einstein equations.\\

In contrast, in an approach based on an ether, it is fully allowed to postulate a relation between absolute effects of gravitation and uniform motion. According to Eq.~(\ref{vecteur_g}), a gravitational field can be characterized by a non-uniform field of the ether pressure $p_e$, or equivalently by a non-uniform field of the ether density $\rho_e$. One then comes naturally to postulate a relation of the kind discussed, when one notes that, in the absence of a gravitational field (thus in SR), a uniformly moving observer would be led to consider that the ``apparent" ether density (for him) is changed, due to the Lorentz contraction. As a result, one is indeed led to characterize the metrical effects of gravity by a strikingly simple relation~\cite{A9}: let $\rho_e^\infty$ be the ether density far enough from any body so that no gravitational field is felt. I assume that at a generic place, thus in the gravitational field, where the ether density is $\rho_e$, material objects are contracted in the ratio
\be \label{coeffbeta}
\beta = \frac{\rho_e}{\rho_e^\infty} < 1,                                   
\ee
and the period $P$ of any clock is dilated in the same ratio, thus becoming $P/\beta$. As for the Lorentz contraction, I assume that the gravitational contraction takes place only in one direction, which can be none other than the direction common to the density and pressure gradients and to the gravity acceleration $\mathbf{g}$. Formally, the dilation of measured distances, resulting from the contraction of measuring rods, occurs as they are compared to distances evaluated with an ``abstract" Euclidean metric, which I assume that the macro-ether M may be equipped with. What I call ``macro-ether" is the three-dimensional space manifold M made of all points whose velocity is the space-averaged velocity of the micro-ether, in short it is the ``absolute space" of the theory. 
\footnote{
~Once we add the time, a such three-dimensional body becomes a ``reference frame": in the case of the macro-ether M, the reference frame is E. The distinction between a body and a frame is a bit subtle. Formally, a frame can be defined as an equivalence class of coordinate systems on space-time, such that each world line with constant space coordinates is the world line of one point in the corresponding body \cite{A16}. 
}
Thus, the preferred reference body M is endowed with two (spatial) metrics: the Euclidean metric $\Mat{g}^0$, and a Riemannian metric $\Mat{g}$. The relation between these two metrics depends only on the field $p_e$, or equivalently on the field $\rho_e$. In the same way, the observers bound to the preferred frame E may use either the ``absolute time" $T$, that which would be measured by a clock which would not be ``affected" by the gravitational field, or the ``local time" $t_\mathbf{x}$, that which is really measured by a clock at point $\mathbf{x} \in \mathrm{M}$.

\subsection{Space-time metric and field equation}

The former assumption about the effects of a gravitational field on rods and clocks is valid for observers bound to the preferred frame E and is easily translated into the form of the physical space-time metric $\Mat{\gamma}$ in coordinates $(x^\mu)$ adapted to that frame, and such that the time coordinate is $x^0 = cT$ with $T$ the absolute time \cite{A9,A15}. The spatial part in the frame E of metric $\Mat{\gamma}$ is the Riemannian metric $\Mat{g}$. The $\gamma_{0i}$ components are zero, which means that the absolute time is a globally synchronized time in the frame E. And, the $\gamma_{00}$ component is the square of the contraction coefficient $\beta$ of Eq.~(\ref{coeffbeta}). Thus:
\be \label{gamma}
\gamma_{00} = f \equiv\beta^2,\quad \gamma_{ij} = - g_{ij},\quad \gamma_{0i} = 0 \qquad (x^0=cT).
\ee
Of course, metric $\Mat{\gamma}$ is a space-time tensor, whose components in an arbitrary reference frame can be obtained by tensorial transformation from the components (\ref{gamma}) valid in the preferred frame E.\\

SR holds true ``locally," first because we have just endowed space-time of a Lorentzian metric which is related in the standard way to actual space and time measurements, and also because the dynamics to be defined in Subsection \ref{ExtensNewton} imposes to the matter particles the upper limit $c$. But, on the other hand, the heuristic interpretation of matter particles as ``localized flows in the micro-ether" sets the ``sound" velocity $c_e$ (Eq.~(\ref{c_e})) as the upper limit. Hence, one must have $c_e=c$ anywhere and at any time, which means that
\be     \label{p_e_rho_e}
p_e = c^2 \rho_e.
\ee
Then, it is found that the wave equation derived for the field $p_e$ by arguments of classical mechanics (see after Eq.~(\ref{c_e})), and hence with absolute time and Euclidean space metric, should hold true in terms of the {\it physical}, curved metric, and with mass replaced by mass-energy~\cite{A9}. This means that I postulate the following equation for the field $p_e$:
\begin{equation} \label{champ1}
\Delta_\Mat{g} p_e - \frac{1}{c^2} \frac{\partial ^2 p_e }{\partial t_{\mathbf{x}}^2}= 4 \pi G \sigma \rho_e,\quad \frac{\partial }{\partial t_{\mathbf{x}}} \equiv \frac{1}{\beta (T,\mathbf{x})}\frac{\partial }{\partial T},
\end{equation}
where $\Delta_\Mat{g}$ is the Laplace(-Beltrami) operator for the curved space metric $\Mat{g}$ and where $\sigma$ is the mass-energy density~\cite{A9}. More precisely, $\sigma$ is
defined as the $T^{00}$ component of the energy-momentum tensor of matter and nongravitational fields {\bf T}, when the time coordinate is $x^0=cT$ with $T$ the absolute time, and in any spatial coordinates that are adapted to the preferred frame E~\cite{A15}. By using the expression (\ref{gamma}) of metric $\Mat{\gamma}$, one finds \cite{A15} that Eq.~(\ref{champ1}) may be rewritten with the flat Laplace operator and the time $T$: 
\begin{equation} \label{champ2}
\Delta f- \frac{1}{f} \left(\frac{f_{,\,0}}{f}\right)_{,0}= \frac{8\pi G}{c^2} \sigma \qquad (x^0 = cT),
\end{equation}
provided one may neglect the time variation of the ``reference ether pressure" $p_e^\infty (T)$. The analysis of cosmological models \cite {A28} shows that this variation takes place over long time scales, of the order of $10^8$ years, hence one may indeed use Eq.~(\ref{champ2}) in the place of Eq.~(\ref{champ1}) in celestial-mechanical applications.\\ 

The SSS solution of Eq.~(\ref{champ1}) leads to Schwarzschild's metric outside the spherical body, but it is not so inside the body~\cite{A9}. For the attraction field $\mathbf{g}$ too, the expression (\ref{vecteur_g}), obtained with arguments of classical mechanics, is assumed to hold true if the gradient operator is now taken to correspond to the curved space metric $\Mat{g}$ (i.e., $(\mathrm{grad} \phi)^i = g^{ij}\phi_{,j}$). In the general static case, the $\mathbf{g}$-field is just the Newtonian attraction field~\cite{A15}.

\subsection{Newton's second law in a curved space-time} \label{ExtensNewton}

In a theory consistent with Lorentz-Poincar\'e relativity, in which space-time is just a mathematical tool, dynamics should be defined by an extension of Newton's second law. Planck's widely used extension of this law to SR (which is equivalent to the formulation with ``longitudinal" and ``transverse" masses, first proposed by Lorentz) applies to a test particle. It consists in inserting the velocity-dependent inertial mass $m(v)$ of SR into the definition of the momentum $\mathbf{P}$ when writing Newton's second law: force = $\dd \mathbf{P}/\dd t$. 

\subsubsection{The case of a constant gravitational field}

It is not too difficult to further extend that dynamics to a test particle {\it in GR}, yet for a {\it constant} gravitational field~\cite{L&L}. To do that, the main problem is to define the {\it time derivative of the momentum}, which is not obvious in a Riemannian manifold, because vectors at different points can not easily be compared. One uses the ``absolute derivative," which is defined in classical textbooks of tensor calculus (e.g. Brillouin~\cite{Brillouin} or Lichnerowicz~\cite{Lichnerowicz}) as a byproduct of the covariant derivative. In fact, Landau \& Lifchitz~\cite{L&L} rewrote the spatial components of the geodesic equation in the given reference frame (in which the gravitational field is constant) so as to make the absolute derivative of $\mathbf{P}$ appear, with respect to the spatial metric in that reference frame. The time parameter on the particle's world line was the proper time of the momentarily coinciding observer in the given frame, thus that time which I call the local time $t_\mathbf{x}$, though synchronized along that world line. The velocity vector was just $v^i \equiv \frac{\dd x^i}{\dd t_\mathbf{x}}$. The components of the force, then, were just what remained on the r.h.s. of the spatial components of the geodesic equation.\\

When I tried to define dynamics in the investigated theory~\cite{A9}, I went in the opposite direction: the gravitational force could be none other than the product $m(v)\mathbf{g}$ (where $v$ is, of course, evaluated with local standards of space and time); as for Landau \& Lifchitz, the momentum had to be $m(v)\mathbf{v}$, and I attempted to define the time derivative of that vector in the three-dimensional space M endowed with the Riemannian metric $\Mat{g}$. Then I realized that, in the general case of a variable gravitational field, the time-dependence of $\Mat{g}$ means that there is actually not a metric on M but instead a one-parameter family of metrics. I thus first thought necessary to restrict myself to the case of a constant gravitational field. I proved that, in the corresponding case of a fixed metric $\Mat{g}$, one is {\it enforced} to use the absolute derivative if one imposes that Leibniz' rule applies to the derivative of a scalar product $\Mat{g}(\mathbf{u}(t),\mathbf{v}(t))$ and that the derivative of a parallel-transported vector must cancel~\cite{A9}. Dynamics of a test particle in a constant gravitational field was thus fixed, and it was also proved~\cite{A9} that, accounting for the postulated form of metrics $\Mat{\gamma}$ and $\Mat{g}$, and of vector $\mathbf{g}$, this dynamics (three scalar equations) {\it implies} that the motion follows the geodesic lines of $\Mat{\gamma}$ (four scalar equations). The result is easily extended to {\it photons}, whose dynamics is also defined, still in the case of a constant gravitational field, by the same extension of Newton's second law (though with $m(v)$ replaced by $h\nu /c^2$ where $h$ is Planck's constant and $\nu$ is the local frequency)~\cite{A18}. By an induction, one might assume that test particles also follow geodesic lines of $\Mat{\gamma}$ in the general case. This is what I did at that time, when I thought unfeasible to define an extension of Newton's second law in the general case~\cite{A9,A18}.
 
\subsubsection{The general case and the energy conservation}

If one adapts the way followed in Ref.~\cite{A9}, and that leads in the ``constant field" case to the absolute derivative, it can be seen that, in the general case where we have a ``time"-dependent vector $\mathbf{u}(t)$ in a manifold M endowed with a family $(\Mat{g}_t)$ of Riemannian metrics, it can still be defined in a natural way a family of ``time" derivatives, say $D_\lambda/Dt$; and that, imposing Leibniz' rule for the derivative of a scalar product $\Mat{g}_t(\mathbf{u}(t),\mathbf{v}(t))$, the parameter must be fixed: $\lambda=\frac{1}{2}$~\cite{A15}. However, this value of $\lambda$ is incompatible with geodesic motion (in view of the postulated form for $\Mat{\gamma}$ and $\mathbf{g}$). Hence, there is a bifurcation in the investigated theory: is Einstein's geodesic motion or Newton's second law to be preferred? The latter is, of course, more consistent with the spirit of Lorentz-Poincar\'e relativity, but the question is principally one of theoretical physics, not one of philosophy. To answer the question, it is necessary to study the problem of the energy in that theory~\cite{A15}. Indeed, the extension of Newton's second law, once fixed, implies some energy equation for a test particle; thus here one gets one energy equation for each given value of $\lambda$. For a dust, that is a continuum made of non-interacting particles, each of which conserves its rest mass, one may apply the extension pointwise in the continuum, thus the dynamics of a dust is deduced from that of test particles. In particular, the energy equation of the constitutive particles gives an energy equation for the continuum. If one rewrites this equation in terms of the energy-momentum tensor $\mathbf{T}$, one has the time component of the flat 4-divergence $\mathrm{div}_0\mathbf{T}$, plus a source term of the form $T^0_0$ times an expression involving only the metric $\Mat{\gamma}$. In order to get a true {\it conservation} equation, this source term must also be a flat 4-divergence. This is possible if, and only if, one fixes that: {\bf i}) the mass-energy density $\sigma$ which is the source of the gravitational field (Eq.~(\ref{champ1})) is indeed the component $T^{00}$ (and not $T^0_0$ or $T_{00}$), {\it and} {\bf ii}) $\lambda=\frac{1}{2}$. Thus, in this theory, {\it one must assume Newton's second law} (in a correct way, i.e., based on a time-derivative that obeys Leibniz' rule). {\it Geodesic motion is recovered only in the particular case of a constant gravitational field.}\\

That extension of Newton's second law can actually be defined in any theory with a curved space-time: if one demands that test particles should follow geodesic lines of the space-metric in the general case, then the form of the gravity acceleration is fixed; it depends on the position and on the {\it velocity} of the particle---as is also the case for the Lorentz force, though here the velocity-dependent part of the force does work~\cite{A16}. 

\subsection{The equation for continuum dynamics} \label{C.dynamics}

An energy equation for a dust can be derived, as we saw, from the extension of Newton's second law assumed for test particles. That energy equation is expressed in terms of tensor $\mathbf{T}$, and it may be assumed to hold true for any continuum: accounting for the mass-energy equivalence, this is the expression of the universality of gravitation~\cite{A15}. To get the complete equation of motion of a continuum, essentially the same technique can be applied. One first computes the expression of the 4-acceleration of a test particle subjected to the assumed dynamics. Considering then a dust and accounting for the relation giving its $\mathbf{T}$ tensor, that expression gives immediately the following equation~\cite{A20}:
\begin{equation} \label{continuum}
T_{\mu;\nu}^{\nu} = b_{\mu},	\qquad b_0(\mathbf{T}) \equiv \frac{1}{2}\,g_{jk,0}\,T^{jk}, \quad b_i(\mathbf{T}) \equiv -\frac{1}{2}\,g_{ik,0}\,T^{0k}.	          \end{equation}
(Indices are raised and lowered with metric $\Mat{\gamma}$, unless mentioned otherwise. Semicolon means covariant differentiation using the Christoffel connection associated with metric $\Mat{\gamma}$.) Equation (\ref{continuum}), including the definition of $b_{\mu}$, is also assumed to hold true for any material continuum. It replaces, in the investigated theory, the equation of GR and other ``metric theories of gravitation," which is
\begin{equation} \label{continuumGR}
T_{\mu;\nu}^{\nu} = 0.	          
\end{equation}

The energy equation previously derived~\cite{A15} is equivalent~\cite{A20} to the  time component of Eq.~(\ref{continuum}). For a dust, Eq.~(\ref{continuum}) {\it implies}, conversely~\cite{A20}, that the constitutive particles obey Newton's second law, and that the rest mass is conserved, i.e.
\begin{equation} \label{massconservation}
\left(\rho^*U^\nu\right)_{;\nu} = 0. 
\end{equation}

\section{Tests and new predictions of the scalar ether theory}

Since this theory predicts Schwarzschild's exterior metric, there is a priori a hope that it may be favourably confronted with observations. To learn more, it is necessary : {\bf i}) to {\it build a consistent ``post-Newtonian" (PN) approximation scheme}, accounting for the preferred-frame effects; {\bf ii}) to develop it until {\it usable formulae} are got; {\bf iii}) to {\it numerically implement} those formulae; and {\bf iv}) to {\it adjust the parameters} in these formulae, using if possible direct observations, in order to check celestial mechanics. As will be shown, points {\bf i}) to {\bf iii}) have been completed quite fully, and point {\bf iv}) partially. There are also easier problems to be investigated.

\subsection{Spherical gravitational collapse in free fall}

A first study of the problem of gravitational collapse ``in free fall" and with spherical symmetry can be done quite easily in this theory~\cite{A18}. According to GR, this situation leads to a very curious catastrophe: a point singularity would form, and this would occur within a finite time of the freely falling particles, but would be beyond the ``temporal horizon" of distant observers~\cite{L&L}. Moreover, in GR, an essentially identical catastrophe should occur, even if one relaxes the simplifying assumptions of spherical symmetry and of a dust (i.e., that the pressure is neglected, whence the free fall). The reason why dust is expected to give the right qualitative response is that, in GR, pressure increases gravity. Thus, one can prove that an SSS equilibrium cannot exist in GR for a perfectly-fluid body that has a too large mass-energy (see e.g. Ref.~\cite{Thirring}). Pressure contributes to gravity also in the investigated scalar theory, for it enters the source $\sigma =T^{00}$ of the gravitational field. Therefore, also in this theory, it suggests itself to begin with the collapse of a dust sphere.\\

Moreover, due to the expression (\ref{vecteur_g}) of the gravity acceleration, the ether density decreases towards the gravitational attraction. According to the idea of a ``constitutive" ether, touched on in subsection \ref{grav=pression}, the matter particles would be merely a kind of localized flows in the fluid micro-ether. Hence, each of them should occupy an increasing volume as the gravitational field increases. Therefore, in a situation of gravitational collapse, there must be a stage from which the ether flows, of which the matter particles are supposed to be made, will ``take all the place" in the fluid (in the domain occupied by the body): from that stage, the macroscopic motion of matter must coincide with the macroscopic motion of the ether, but the latter defines the privileged reference frame.\\

The study of collapse is hence made with the additional assumption that the motion of the collapsing body is bound to the preferred frame E. (In other words, the collapsing body is in fact at rest in E, hence in particular its radius, as evaluated with the invariable Euclidean metric $\Mat{g}^0$, is a constant $R$! The ``implosion" means that, as evaluated with the {\it physical} space metric $\Mat{g}$, which depends on time, the radius $R'$ of the body decreases.) Under these assumptions, the collapse can be followed analytically~\cite{A18}. The salient results are: {\bf i}) These conditions preclude that one starts from a static situation; said differently, the envisaged model of a collapsing body can apply only after a more complex stage of rearranging. {\bf ii}) The density of the collapsing body is necessarily {\it uniform} during that stage. {\bf iii}) The radius $R'$, evaluated with the physical metric $\Mat{g}$, passes a minimum. In other words, after some time, {\it the implosion becomes an explosion}, which means that there can be no point singularity.
\footnote{
~The study in Ref.~\cite{A18} was made with the assumptions of a geodesic motion (instead of assuming Eq.~(\ref{continuum})), and the source of the gravitational field was assumed to be $T^0_0$, not $T^{00}$ (see Eq. (67) in Ref.~\cite{A18}). However, the study remains analytical with the new assumptions, and the same qualitative results {\bf i}) to {\bf iii}) are obtained. (Unpublished.)
}

\subsection{Gravitational effects on light rays} \label{rayons}

In order to test the investigated theory, it has been built an approximation scheme that applies to a system of extended bodies with weak and slowly varying gravitational field, thus a PN scheme. That scheme~\cite{A19,A23} differs from the ``standard" PN scheme developed by Fock~\cite{Fock59} and Chandrasekhar~\cite{Chandra65} in one important point: in the former~\cite{A19,A23}, all fields are expanded with respect to the small parameter. In the standard scheme~\cite{Fock59,Chandra65}, the gravitational field is expanded, but the matter fields (pressure, density, velocity) are not. The adimensional small parameter is classical~\cite{Fock59,MTW,Will}:
\be         \label{epsilon}
\varepsilon \equiv (U_{\mathrm{max}}/c^2)^{1/2 },
\ee
where $U\geq 0$ is the Newtonian potential and $U_{\mathrm{max}}$ is its maximum value in the system. (One may ask what is the Newtonian potential in another theory as Newton's! This detail will be answered in Subsection \ref{PNA} below.) The orbital velocities are at most of the order $\varepsilon .c$, and this actually implies that the gravitational field is slowly varying. This is accounted for by assuming that the time derivatives of the relevant fields are one order higher in $\varepsilon $ than the corresponding space derivatives. The practice of this scheme consists {\it precisely} in writing Taylor expansions with respect to $1/c^2$, while considering $x'^0\equiv T$, not $x^0\equiv cT$, as the relevant time variable for the expansions,
\footnote{
~Note that the choice between $x^0$ and $x'^0$ does matter, since $(x'^0/x^0)^2=1/c^2$, the actual small parameter in the expansions.
}
and in admitting that all relevant fields are order zero in $1/c^2$. This way of doing can be partly justified by taking units which are such that $U_{\mathrm{max}}=1$ (cf. Eq.~(\ref{epsilon})), but a more satisfying justification will be provided in Subsection \ref{PNA} below. (As to the standard PN scheme~\cite{Fock59,Chandra65}, its practice is just the same, except for the fact that the matter fields are not expanded.) For a preferred-frame theory like the theory investigated, one has also to admit that the ``absolute" velocity of the mass center of the gravitating system is at most of the order $\varepsilon .c$, too, and this is entirely compatible with the relevant orders of magnitude.\\

This scheme has first been applied~\cite{A19} to evaluate the preferred-frame effects that may affect the photons at the first PN (1PN) approximation, those being seen as light-like particles subjected to the extension of Newton's second law defined in Section \ref{ExtensNewton}. (The 1PN approximation is the second iteration of the PN scheme, the first one or 0PN approximation being Newton's theory.) To do so, the 1PN expansions of the scalar field equation, the metric $\Mat{\gamma}$, and of the equation of motion for the photon, have been written. It turns out that the 1PN equation of motion for a photon, which is obtained thus, is identical to the 1PN expansion of the {\it geodesic} equation of motion for a light-like particle in the metric $\Mat{\gamma}$. Hence, it merely remains to compute the relevant terms of $\Mat{\gamma}$ in the reference frame bound to the mass center of the system. This has been done by using a Lorentz transform (relative to the flat metric $\Mat{\gamma}^0$) of $\Mat{\gamma}$, taking into account the 1PN expansion of the latter in the preferred frame. When the system reduces to a unique massive body with spherical symmetry (which is the assumption used when evaluating the gravitational effects on light rays in GR), the relevant terms in metric $\Mat{\gamma}$ are the same as for the 1PN expansion of Schwarzschild's metric. Therefore, {\it there is no preferred-frame effect for photons at the 1PN approximation, and the effects predicted by this scalar theory are the same as in GR.}

\subsection{Matter creation/destruction in a variable gravitational field}

One could a priori expect that, in a ``relativistic" theory of gravitation involving the mass-energy equivalence, the rest-mass should not be conserved in general (since this is already the case in SR, in agreement with experiments in particle physics), and that, more specifically, matter might be produced or destroyed by an exchange with the gravitational energy, in a variable gravitational field. However, the dynamical equation usually stated in relativistic theories of gravitation, i.e., $T_{\mu;\nu}^{\nu} = 0$ (which, in GR, is a consequence of the Einstein equations), implies that the rest-mass is exactly conserved for an isentropic perfect fluid. This has been proved by Chandrasekhar~\cite{Chandra69}.

By adapting his reasoning, one may show that things go differently in the scalar theory: according to the latter, matter may indeed be created or destroyed for an isentropic perfect fluid in a variable gravitational field. The creation/destruction rate $\hat{\rho}$ is proportional to the pressure $p$ in the fluid and to the logarithmic variation rate of the gravitational field~\cite{A20}. It comes from the different dynamical equation which applies in this theory, Eq.~(\ref{continuum}), but the real reason is that a true {\it conservation} equation applies to the (material plus gravitational) energy in the scalar theory: exact conservation of mass and energy are, in general, incompatible---except for the case of a dust. For a weak field, it is found that 
\be \label{rho_chapeau}
\hat{\rho}=\frac{p}{c^4}\frac{\partial U}{\partial T},
\ee
where the time derivative of the Newtonian potential has to be taken in the preferred frame and hence depends on the ``absolute velocity" of the body which is considered. For the expected velocities, which are of the order of $10^{-3}c$ at most, one finds that the relative production rate $\hat{\rho}/\rho$ would not exceed $10^{-23}\,\mathrm{s}^{-1}$ near the surface of the Earth (the time-averaged rate being lower due to the Earth rotation). This seems to be small enough to explain why it has not been detected so far. Nevertheless, the rates would be much higher inside compact and very massive objects, in fact already inside the Sun.

The derivation of this result~\cite{A20} is not limited to the case of an isentropic perfect fluid, instead it starts from a general type of perfect fluid, which may be used to phenomenologically describe some entropy production due to dissipative processes (although such processes cannot take place in a ``truly perfect" fluid). However, in this case where the rest-mass is not conserved, there is some ambiguity regarding the formulation of the second law of thermodynamics. First, we may consider that the entropy of a small volume of fluid, which is followed in its motion, is $\delta S=\sigma. \delta m$ with $\delta m$ the (variable) rest mass of this fluid element and $\sigma$ a ``specific entropy"---which, by definition, should be conserved for an adiabatic process. In that case, mass creation, by itself, makes entropy increase, so that the second law, when applied to an ``adiabatic" process, forbids matter {\it destruction}~\cite{Prigogine}. But this formulation of the second law of thermodynamics allows that $\sigma $ decreases, provided that there is simultaneously enough matter creation. This means that usual thermodynamics does not apply: it would be possible to extract heat from a cold source if, simultaneously, enough matter would be created. Therefore, it has been given a different formulation of the second law, which essentially says that usual thermodynamics does apply, and which, consequently, allows a {\it reversible} production or destruction of matter---as is predicted by the scalar theory in a variable gravitational field~\cite{A20}. 

\subsection{Homogeneous cosmological models and accelerated expansion}

The scalar theory postulates a gravitational contraction of material bodies (Subsection \ref{Metric}). This contraction (in the direction $\mathbf{g}$ only) defines, at a given time $T$, the relationship that exists between the two spatial metrics on the preferred reference body M: the Euclidean metric $\Mat{g}^0$, which makes M a rigid body, and the physical, Riemannian metric $\Mat{g}$. It is hence natural to complete this by introducing a scale factor $R(T)$, which leaves to that relation the possibility to depend on the time $T$. It is even necessary to do so, for the conservation of ether implies that $R(T)$ is determined by the field $\rho _e$ and depends indeed on $T$~\cite{A28,B20}, hence the conservation of ether (which plays a crucial role~\cite{A8} in finding the equation for the field $p_e$) would not hold if one would omit the $R(T)$ factor. When there is expansion in the usual sense, i.e., when $R(T)$ increases with $T$, this means that there is a ``cosmological" {\it contraction} of bodies, with respect to the Euclidean metric $\Mat{g}^0$. From the analogy with the metrical effects of a uniform motion and of a pure gravitational field, it is then natural that one leaves also the possibility of a ``cosmological" time-dilation. This is done by introducing an exponent $n$ ($n=0$ if there is no cosmological time-dilation).\\

The foregoing applies to a heterogeneous universe, as we can see it. (From what we know about the spatial variations of the density at a very macroscopic, astronomical scale, our real Universe is {\it extremely} heterogeneous.) In the scalar theory, the cosmological principle can be formulated in a logically sparing way: in order that the geometry defined by metric $\Mat{g}$ be (spatially) homogeneous, it is necessary and sufficient that the field $p_e$ depend only on the time $T$ (and hence that $p_e(\mathbf{x},T)=p_e^{\infty}(T)$ for any $\mathbf{x}$ and at any time $T$). This implies that the space (M, $\Mat{g}$) is also isotropic, for $\Mat{g}$ is then equal to the Euclidean metric $\Mat{g}^0$, up to the factor $R(T)$. This implies also that the attraction field (\ref{vecteur_g}) cancels. The local time $t_{\mathbf{x}}$ becomes then a uniform ``cosmic time" $\tau$. When expressed in terms of $\tau$, the space-time metric $\Mat{\gamma}$ is then the Robertson-Walker metric with zero spatial curvature. The equation (\ref{champ1}) for the field $p_e$ implies immediately that matter also must be homogeneously distributed, and that equation can be rewritten in the form of a ``Friedmann equation" which holds true independently of any assumption on the material behaviour. The latter equation implies that {\it the cosmic expansion is necessarily accelerated}, according to that theory~\cite{A28,B20}. This is independent also on the assumption of a cosmological time-dilation. The interpretation of redshifts observed on distant supernovae has precisely led to the conclusion that the cosmic expansion is accelerated~\cite{Riess,Perlmutter}.\\

The spatial components of the dynamical equation (\ref{continuum}) reduce to the conservation of momentum in that expanding (or contracting) universe, and the time component expresses the conservation of energy. Using the latter, the equation for the scalar field reduces to an ordinary differential equation for the energy density $\epsilon $, that depends only on the time-dilation exponent $n$, and that admits an analytical solution for any value of $n$. There are three possible scenarios~\cite{A28,B20}: {\bf i}) expansion only, the density $\epsilon $ tends towards infinity as the cosmic time $\tau $ goes back to $-\infty $; {\bf ii}) one contraction-expansion cycle with a bounce at a finite density; {\bf iii}) an infinity of non-necessarily identical such cycles. The latter scenario is the most acceptable physically, for the other two involve the existence of an ``end of times" after which the theory cannot say anything. In all cases, the phases with (accelerated) expansion end with an {\it infinite dilution} ($\epsilon =0$) that is reached within a {\it finite} cosmic time. For scenarios {\bf ii}) and {\bf iii}), the maximum density $\epsilon _{\mathrm{max}}$, reached in the contraction phase that precedes the currently observed expansion, is not determined from the knowledge of the current values of the density and the Hubble parameter (one would have to know also the current value of the deceleration parameter $q$). If ``cosmological" redshifts at $z\simeq 4$ are indeed observed, it means that $\epsilon _{\mathrm{max}}$ must be at least hundred times the current density. In that case, the cosmic time elapsed since $\epsilon _{\mathrm{max}}$ was reached is at least equal to several hundreds of billions of years. Thus, in this cosmology, the time scale is very large. (By the way, this implies that $p_e^{\infty}(T)$ may indeed be considered constant in celestial mechanics.)

\subsection{Asymptotic post-Newtonian approximation} \label{PNA}
\subsubsection{Definition of the approximation scheme}

For the study of gravitational effects on light rays~\cite{A19}, summarized in Subsection~\ref{rayons}, the operational basis of a PN approximation scheme had been set, which scheme was quite similar to the standard scheme~\cite{Fock59,Chandra65} used in GR (up to the difference mentioned, about the expansion of the matter fields). However, the general meaning of that scheme remained rather obscure, and things became even less clear when it was tried to study the effects of gravitational radiation: in that case, it was indeed necessary to consider, not any more $x'^0\equiv T$, but instead $x^0\equiv cT$, as the relevant time variable for the expansions, so as to obtain a wave equation for the  gravitational potential. (The general meaning of the standard scheme is even more obscure, for it is very difficult to understand why the matter fields should not have to be expanded if asymptotic expansions in the usual mathematical sense are sought for.) In order to build a consistent asymptotic scheme, thus one involving a small parameter that one really can make tend towards zero, it was searched how to define a {\it family} ($\mathrm{S}_\varepsilon$) of perfect-fluid systems of massive bodies, indexed by the classical field-strengh parameter $\varepsilon$, and which really would justify the formal expansions in $1/c$~\cite{B18,A23}. 
\footnote{
~The assumption that each body is made of a perfect fluid, i.e., a material with a spherical stress tensor, is not a strong restriction in practice, because the non-spherical part of the stress is unlikely to play any significant role in PN gravity~\cite{MTW}. One could easily adopt a more general material behaviour---at the price of redoing (all) PN computations. But it is necessary to assume a definite state equation, in order to have a closed system of equations.
}\\

Because the Newtonian potential, $U=GM/r$ for a spherical body, is by the definition (\ref{epsilon}) like $\varepsilon ^2$, one is led to admit that the velocity field $\mathbf{u}$ is like $\varepsilon $ (cf. the case of a circular orbit, where $u^2=GM/r$). Imposing the condition that the spatial dimensions should be of the same order of magnitude for all members of the family ($\mathrm{S}_\varepsilon$), i.e., should be like $\varepsilon ^0$, the condition $U=\mathrm{ord}(\varepsilon ^2)$ imposes that the masses of the bodies are like $\varepsilon ^2$ also, hence the same must be true for the density field $\rho $. Then, if one changes the time and mass units thus for system $\mathrm{S}_\varepsilon$: $[\mathrm{T}]_\varepsilon = [\mathrm{T}] /\varepsilon $, $[\mathrm{M}]_\varepsilon = [\mathrm{M}].\varepsilon^2 $, the fields $U$, $\rho$, and $\mathbf{u}$, become already order 0 in $\varepsilon $, moreover the small parameter $\varepsilon $ becomes proportional to $1/c$ ($c$ being evaluated in the new units). In order that also the pressure field $p$ be $\mathrm{ord}(\varepsilon^0) $ in the new units, and this with a state equation which should be independent of $\varepsilon $ (in the new units), it must be that in fact (i.e., in invariable units), $p$ is like $\varepsilon ^4$ (hence the state equation actually depends on $\varepsilon $ in invariable units). It was then  realized~\cite{A23} that, if one starts from a set of fields $U$, $\rho $, $p$, and $\mathbf{u}$, that obeys the Newtonian equations, and if one applies to this set a natural $\varepsilon $-dependent transformation defining new fields having  the orders just defined, these new fields are still exact solutions of this same equations. (Since the velocity is like $\varepsilon $, the characteristic time is like $1/\varepsilon $, and this of course must be accounted for in the definition of the new fields.) One thus obtains the ``weak-field limit" of Newtonian gravity itself, which is hence trivial: in the $\varepsilon $-dependent units, the fields are simply independent of the small parameter $\varepsilon $. \\

This exact similarity transformation in Newtonian gravity (NG) was also noted by Futamase \& Schutz~\cite{FutaSchutz}. It obviously does not hold true for the scalar theory (nor for GR), since the equations are complicated by the presence of the curved metric and its dependence on the scalar gravitational field. But a family of systems can be and has to be defined by a family of initial conditions, because the natural boundary-value problem in that theory is the full initial-value problem (due to the fact that Eq.~(\ref{champ2}) is hyperbolic)~\cite{A23}. It then seems obvious how to define the Newtonian limit of the scalar theory: {\it one applies the Newtonian similarity transformation to the initial data}. The matter fields are indeed common to NG and to a ``relativistic" theory of gravitation like the scalar theory, up to small modifications. It remains to define an equivalent of the Newtonian potential, but this turns out to be possible and easy for this scalar theory. The Newtonian limit of a ``relativistic" theory has to impose that the curved metric tends towards a flat metric as $\varepsilon \rightarrow 0$. One thus finds that, for this scalar theory, the field
\be
V\equiv \frac{c^2}{2}(1-f)
\ee
is indeed a natural equivalent of the Newtonian potential~\cite{A23}. In the $\varepsilon $-dependent units, all relevant fields are $\mathrm{ord}(\varepsilon^0) $ (at the initial time, hence also in its neighborhood), moreover the small parameter $\varepsilon$ is proportional to $1/c$. Because only $1/c^2$ enters the equations, Taylor expansions in $1/c^2$ are postulated (at least until some order; in fact, only the first-order in $1/c^2$, or 1PN approximation, has been studied in that theory). In this ``asymptotic" PN scheme, it is important to expand also the {\it initial conditions.} The 0-order expanded fields obey just the equations of NG, hence there is a correct Newtonian limit in the scalar theory~\cite{A23}.\\

Having built that scheme, I learned that a scheme of the same kind had been proposed for GR by Futamase \& Schutz~\cite{FutaSchutz}. In my scheme, one starts from a very general self-gravitating system with a spatially bounded distribution of matter, whereas the initial condition assumed for the spatial metric and its time derivative by Futamase \& Schutz~\cite{FutaSchutz} is special in that it enforces spatial isotropy. 
\footnote{
~This is connected with the fact that, in GR, precisely that part of the initial data is subjected to four nonlinear ``constraint equations" (which are just the time-time and time-space components of the Einstein equations)~\cite{Stephani82,Rendall00}. It might therefore turn out to be difficult to build an ``asymptotic" scheme in GR for general initial conditions and/or in other gauges.
}
The change of units, that allows to justify the formal expansions in $1/c^2$, does not appear in Ref.~\cite{FutaSchutz}. Lastly, their scheme is not precisely compared with the ``standard" PN scheme~\cite{Fock59,Chandra65}, which consists in expanding the gravitational field, but not the matter fields, in powers of $1/c^2$. For the scalar theory, it has been proved that the standard scheme is incompatible with this interpretation of the $1/c^2$ expansions as asymptotic expansions~\cite{A23,O1}.

\subsubsection{Equations of motion of the mass centers (EMMC's)} \label{EMMC}

In a relativistic theory of gravitation, any form of material energy must both contribute to the gravitational field and be subjected to its action. It is not obvious, therefore, to state which energy density may be used so as to define a relevant mass center. Two arguments may justify the choice made, of the rest-mass density (in the preferred frame), denoted by $\rho$~\cite{A25}: {\bf i}) this density, or rather its 1PN approximation, obeys the usual continuity equation, which allows to commute time differentiation and barycentration; {\bf ii}) rest-mass is well-correlated with astronomical observations, for it is indeed the presence of matter in the usual sense, thus characterized by its rest-mass density, that leads to the electromagnetic emission detected by the telescope. If one adds to $\rho$ some kinetic and potential energy, as is done in the literature on PN calculations in GR~\cite{Fock59,Chandra65,Chandra69,Weinberg,MTW,Will}, then one looses criterion {\bf i}) and one weakens the validity of criterion {\bf ii}), although the final EMMC's may be simpler. The (1PN) EMMC's are got by integrating in the volume of the different bodies the spatial components of the local 1PN equations of motion. In the asymptotic scheme used, identifying the powers in $1/c^2$ leads to {\it separate} the local equations into {\it exact} equations for the orders 0 and 1 in $1/c^2$, which does not occur in the standard scheme~\cite{A23,O1}. Hence, the same occurs for the EMMC's~\cite{A25}, whose structure is thus quite different in the two schemes. In particular, a number of integrals of the 0-order (Newtonian) fields $U\equiv V_0$, $\rho _0$, $p_0$, and $\mathbf{u}_0$, enter the EMMC's for the 1PN corrections (the equations of the order 1), which depend thus on the {\it internal structure} of the bodies and on their {\it internal motion}. It is hard to see how this influence might be artificially cancelled, and this result seems true independently of the theory considered.\\

In order to get usable EMMC's, two simplifying facts may be used, because they do occur e.g. for the solar system: {\bf i}) the main bodies are (nearly) spherical, and {\bf ii}) the bodies are well-separated. In order to account for point {\bf i}), it is assumed, merely at the stage of calculating the 1PN {\it corrections}, that the Newtonian density $\rho _0$, hence also the pressure $p_0$ and the Newtonian self-potential, are spherical. The good separation allows to introduce a small parameter $\eta$, which is the maximum of the ratio between the radius of a body and its distance to an other body. To correctly take this into account, it is hence again necessary to introduce a family of systems, this time a family of 1PN systems. (Since the equations of the asymptotic 1PN approximation are exact equations that make a closed system, one may here forget the field-strength parameter $\varepsilon$.) In a first work~\cite{A26}, it had been thought sufficient to consider the parameter $\eta $ without introducing such family. But then, it was only possible to compare certain couples of terms in the EMMC's, without possibly assigning a definite order in $\eta$ to an individual term, and it was decided to retain only the term of the lowest order in those couples. This led to neglect in the final equations some terms which turn out to be numerically significant~\cite{A32}. In GR, the method used to account for the good separation is different: it consists in retaining multipoles up to a certain order~\cite{Papapetrou,DSX91,DSX92}. I think that an asymptotic expansion with respect to $\eta $, as is done here, brings us closer to a numerical control of the error involved in neglecting some terms. The final EMMC's depend on three {\it structure parameters} per each body, and also on the spin velocity of each body~\cite{A32}. 

\subsubsection{Point-particle limit and violation of the weak equivalence principle} \label{point-particle}

Once an influence of the internal structure on the motion of the mass centers had been found (and this already at the 1PN approximation), it became interesting to check whether a such influence might survive in the limit of a point particle. This would constitute a patent violation of the ``weak equivalence principle" (WEP), according to which the acceleration of the mass center of a body in a gravitational field is the same for any kind of body, at least for very small bodies. The WEP applies undoubtedly to Newton's theory, but it is not so obvious, after all, that it should hold true in a relativistic theory of gravitation---unless one restricts the discussion to {\it test particles}. By definition, such (point) particles do not influence the gravitational field, and for that reason most theories (including the investigated theory) predict that all have indeed the same acceleration. But in fact a massive body, however small, does influence the gravitational field: even in NG, this ``self-field" is not always negligible inside that body itself. If it turns out to give no net contribution to the motion of the body's mass center, this is due to the actio-reactio principle, which applies in NG. But the latter principle can not even be formulated in a nonlinear theory (see the footnote in Subsection \ref{podlaha}), hence in any such theory it is a priori possible that the influence of the self-field survive in the point-particle limit.\\

In order to define rigorously that limit, it is once more appropriate to introduce a family of 1PN systems, that are identical up to the size of the body numbered (1), which size is precisely the small parameter $\xi$~\cite{A33}. This allows one to evaluate the limits as $\xi \rightarrow 0$ of the different terms in the 1PN correction to the acceleration of the mass center of the small body. One of those terms is the quotient by the mass of the small body, of the integral, within that body, of
\be \label{K4}
f_i = \frac{1}{2} \rho_0 u_0^j u_0^k (h_{jk,i} - h_{ik,j})
\ee
(where $\Mat{h}$ is the non-Euclidean part of the 1PN spatial metric). This term comes from the term $\Gamma ^i_{jk}T^{jk}$ (with $\Gamma ^\mu _{\nu \rho }$ the Christoffel symbols of the space-time metric $\Mat{\gamma }$) in the spatial component ($\mu =i$) of the dynamical equation $T^{\mu \nu}_{;\nu} = b^{\mu}$ [cf. Eq. (\ref{continuum})], which term itself arises from the covariant derivative $T^{i \nu} _{;\nu}$. Therefore, the term (\ref{K4}) occurs also in GR. In the scalar theory [in the version discussed here], $\Mat{h}$ depends on the spatial derivatives $U_{,i}$ of the Newtonian potential, hence $f_i$ includes the {\it second} spatial derivatives of $U$. Now the self part of those second derivatives is order 0 in $\xi$, and it depends on the internal structure of the small body. For this reason, the PN acceleration of the small body does depend on its structure even at the point-particle limit in that theory. This cannot happen when the spatial metric is conformal to an Euclidean metric---as is the case at the 1PN approximation for GR {\it in the harmonic gauge}, at least with the standard PN scheme~\cite{Fock59,Chandra65}, as also with the asymptotic scheme of Futamase \& Schutz \cite{FutaSchutz}, owing to the initial condition used in the latter \cite{GAS04}. However, the {\it standard} (as opposed, here, to ``harmonic") form of Schwarzschild's metric, thus the standard SSS solution of GR, exhibits just the same dependence on the derivatives $U_{,i}$ as does the spatial metric of the scalar theory. (With, in that case, $U=GM/r$.) Hence, in a gauge in which the SSS solution is this standard metric, 
\footnote{
~The four gauge equations $\gamma_{0i}=0$, $\gamma_{00}(2+\gamma_{jj})=-1$, do the job. The general SSS metric may be written thus in ``Cartesian" coordinates $x^1\equiv r\sin\theta\cos\phi$, $x^2\equiv r\sin\theta\sin\phi$, $x^3\equiv r\cos\theta $: $\gamma_{00}=p(r)$, $\gamma_{0i}=b(r)\frac{x^i}{r}$, $\gamma_{ij}=-\frac{1}{r^2}\left[a^2(r)\delta_{ij}+\left(q(r)-\frac{a^2(r)}{r^2}\right)x^ix^j\right]$. The Einstein equations impose $p(r)=1-2\frac{m}{a(r)}$, $q(r)=\frac{a'^2(r)-b^2(r)}{p(r)}$~\cite{Brumberg}. My gauge equations give $b=0$, $q=\frac{a'^2}{p}$, $a'^2= 2(1-\frac{2m}{a})(1-\frac{a^2}{r^2}) +1 $, thus excluding the harmonic solution $a(r)=r+m$. The Newtonian limit imposes $a(r)\sim r$ as $r\rightarrow \infty$. A solution of the last differential equation such that $a(r) \ne r$ would then have to cross the line $a(r) = r$, which is impossible by uniqueness. With $a(r) = r$, we have the standard Schwarzschild metric. 
}
the 1PN metric of the generic case should still contain a dependence on the $U_{,i}$'s, hence might plausibly lead to a violation of the WEP, too. Thus, it might be the case that, depending on the gauge, the WEP could be violated at the 1PN approximation of GR, or not---but this would be hard to check.\\

The order of magnitude of the violation of the WEP in the scalar theory has been assessed: it is dangerous, but it may remain compatible with the data of the experiments that aim at testing that principle. I note that the analysis of these experiments (mainly based on a torsion balance, see Fischbach \& Talmadge~\cite{Fischbach} and references therein) is purely Newtonian. I find this quite surprising in view of the magnitude of the PN corrections~\cite{A33} and the extreme precision (quite amazing in fact), of $10^{-12}$, which is announced using such mechanics experiments.

\subsection{Celestial mechanics in the solar system}

\subsubsection{Remarks on the test of an alternative theory} \label{test-alternative}

To test an alternative theory in celestial mechanics is obviously much more than just add PN corrections calculated from an ideal (SSS) situation to the Newtonian ephemeris (cf. Subsect. \ref{podlaha}). It cannot even reduce to numerically implement the general EMMC's of the theory, taking the values of the parameters in the astrodynamical literature, and to compare the result of the integration with a reference ephemeris. This task, already much more complex, indeed cannot be enough, for two reasons:\\

{\bf i}) {\it The astrodynamical parameters (masses, initial conditions, etc.) must be regarded as adjustable parameters, whose optimal values depend on the theory}~\cite{A26}. (The initial conditions depend even, within a given theory, of the precise model which is considered, thus on the list of the bodies taken into account, the integration time interval, etc.~\cite{Bretagnon,A30}.) In addition, a new theory usually depends on a different set of parameters, as compared with the reference theory. For instance, in the investigated theory, there is one additional vector parameter, namely the ``absolute velocity" $\mathbf{V}$ of the zero-order mass center of the solar system. (This is a constant vector as far as the system is considered isolated.) Moreover, the ``asymptotic" PN scheme which is used leads to separated equations for the orders 0 and 1, which has consequences on the numerical integration (e.g. one needs to implement a ``reinitialization" procedure~\cite{A29}), hence on the optimal values of the parameters. Further, the asymptotic scheme makes already the 1PN approximation depend on some {\it structure parameters} of the bodies. For the reasons summarized in point (\ref{EMMC}) (see Ref.~\cite{A32} for more details), I think that this dependence is inherent in all relativistic theories of gravitation, {\it including GR.}\\

{\bf ii}) A reference ephemeris (e.g.~\cite{Newhall83,Standish95,Standish98}) is the result of a final integration following an adjustment of constants (which adjustment itself is based on an iteration of the integration procedure~\cite{Newhall83}), this being done in the currently accepted framework, i.e., the reference theory (GR), {\it plus} the accepted approximation scheme---thus the standard Fock-Chandrasekhar scheme leading, after simplifying assumptions, to the Einstein-Infeld-Hoffmann equations. The adjusted constants include~\cite{Standish90} the astrodynamical parameters that enter the EMMC's, which parameters cannot be observed directly. But they do not reduce to the former: the analysis of the {\it basic observations} themselves (transit, photographic, and CCD observations, radar ranging, VLBI data) depends on a plenty of not-accurately-known parameters (e.g. for optical observations: phase corrections, catalogue drifts,...), which are thus adjusted for the construction of the ephemeris~\cite{Standish90}. This means simply that {\it the observations are biased by the reference theory} in some way, which is difficult to quantify. Also observations which are analysed fully independently of the ephemeris construction are biased by the theory, because the latter is used for this analysis, anyway.\\

As long as these two points are not carefully accounted for (and this is admittedly a hard work), one will stay in a situation where any theory, which is not an extension of the reference theory (GR), and/or which just uses a {\it different approximation scheme}, will be rejected, and where such extensions might be accepted only for values of the additional constants which make the ``new theory" coincide for all practical purposes with GR. 

\subsubsection{Parameter adjustment and first results}

The analysis of observations uses very specialized knowledge and one may hope, at least in a first step, that the theory/observation coupling noted at point {\bf ii}) above does not affect too seriously the results. Therefore, until now, I limited myself to account for point {\bf i}). That is, I studied the adjustment of the free parameters entering the EMMC's of the scalar theory (namely, the masses and the initial conditions, plus vector $\mathbf{V}$) on a reference ephemeris. To begin with, an adjustment program has been built, thus a program for parameter optimization in celestial mechanics, based on the Gauss algorithm plus a directional minimization, and it has been tested with the Newtonian equations~\cite{A31}. By adjusting the latter ones using data taken from the DE403 ephemeris of the Jet Propulsion Laboratory~\cite{Standish95}, it was possible to reexamine the question of Mercury's perihelion with current data and with the power of computers. (The relativistic ephemerides such as DE403 are all based on the so-called Einstein-Infeld-Hoffmann equations---hereafter {\bf EIH}, although these equations were actually derived, prior to these authors, by Lorentz and Droste.) The result of this adjustment is that, using the Newtonian equations, the standard-GR ephemeris~\cite{Standish95} cannot be reproduced with a smaller angular error than $20''$ per century. Moreover, the ``residual" advance in the longitude of the perihelion of Mercury (as described by DE403, and with respect to a Newtonian calculation) is indeed $43''$ per century. In addition, the adjustment of the initial conditions is essential~\cite{A31}. This adjustment program was then used to minimize over 3000 years the difference between the DE403 ephemeris, that is based on standard-GR equations for $N$ interacting bodies and takes into account a lot of minor bodies, and Newtonian calculations considering only the main bodies, optionally augmented with the ``Schwarzschild corrections" (one then obtains much smaller differences)~\cite{A30}.\\

Let me come to the adjustment of the 1PN EMMC's obtained~\cite{A32} with the asymptotic PN scheme of the scalar theory, in short ``my equations". It uses, as it must, the reinitialization procedure~\cite{A29} mentioned above, which unfortunately increases significantly the computer time and limits the accuracy. It also uses Lorentz transforms of the flat metric, in order to go from the preferred reference frame to the solar-barycentric frame and vice-versa. For quite a time, it was found very difficult to optimize the masses, whereas the optimal masses are certainly significantly different from the standard-GR masses, due to the difference in the EMMC's. (This was due to an algorithmic defect: ``perturbatively," one should not need to use the current trial for the new masses when calculating the PN corrections, because the mass modifications themselves have the status of PN corrections; nevertheless, this turns out to be necessary for the optimization algorithm.) Now one of the most important differences between the EIH equations and ``my equations" is that, in the latter ones, the spin (the self-rotation) $\mathbf{\Omega}$ of a body influences the 1PN acceleration of {\it that body} (thus a ``self" term), in first order in $\mathbf{\Omega}$. This is not specific of the scalar theory: instead, as for the violation of the WEP (point \ref{point-particle}), the same should be true in GR in some gauges, though not in the harmonic gauge.
\footnote{\ 
Since the first version of this contribution was written, the asymptotic PN scheme has been investigated in detail for GR in the harmonic gauge \cite{A36}. It has thus been found that the asymptotic PN scheme leads to add to the EIH equations a ``self" term. That term is quadratic in $\mathbf{\Omega}$, but it depends on the internal structure, and it does not seem to be negligible for the giant planets (see gr-qc/0504016v1, p. 28).
} \label{!!!}
The terms that are linear in $\mathbf{\Omega}$ have a significant 
numerical effect (although this effect is close to periodic). Hence, without an efficient mass optimization, the trajectories would be quite different from those predicted by the reference ephemeris. Therefore, in a first step, the spin rates were set to zero. Then it could be obtained differences with the DE403 ephemeris at the arc-second-per-century level (at most $3''7$, found for the longitude of Mercury), while leaving the masses of the bodies at their standard values but for those of the Sun and Jupiter~\cite{O1,B21}. Recently, it has been succeeded in optimizing masses in a more efficient way, hence it has been began to adjust the equations using the spin vectors as given~\cite{BDL} in the astrodynamical literature (these values also may be partly affected by a theory-dependent adjustment). The result that seems to emerge is that the optimization algorithm leads to very important mass modifications (a few percents, except for Neptune, for which the mass would increase by more than 50 percent, and for Mercury's mass, which was kept fixed); and that, using these optimized masses, the differences are at the level of $10''/\mathrm{cy}$. As to the ``absolute" velocity $\mathbf{V}$ of the solar-system barycenter, which is also adjusted, it was found to be 3 km/s without the spins~\cite{O1,B21}, and it is currently found to be less (300 m/s) with them. It should be realized that, the solar system being assumed isolated here, one cannot hope to precisely assess $\mathbf{V}$, because the global motion of the solar system is influenced by neighbouring stars and even by the whole Galaxy. It must also be emphasized that these results are not definitive ones (because the numerical process is noised by the reinitializations and, more generally, is improvable), and that here the comparison basis is with a standard-GR ephemeris---thus itself a fitting of observations by the EIH equations, all observations being analysed within standard GR (see point \ref{test-alternative}). Hence, the error found here is certainly much higher than that which would be found by starting from primary observations.

\subsection{Energy loss by gravitational radiation}

In GR, most studies of gravitational radiation and its effect on the period of a binary system start from a linearization of the equations~\cite{L&L,Weinberg,MTW}. In the harmonic gauge, that linearization leads to the well-known ``quadrupole formula of GR". It is not difficult to make a such linearization in the investigated scalar theory, and thus to obtain the equivalent of that formula. However, we are here in a domain where the accuracy of the observations (of the timing of the pulses emitted by binary pulsars) is very high, and so also is the precision of the comparison theory/ observation~\cite{TaylorWeisberg}. Therefore, it is desirable that the link between the equations of the theory and its numerical predictions should be obtained with the help of a well-defined approximation scheme. One possibility would be to make a complete analysis of the relativistic 2-body problem up to the order $1/c^5$ or higher. Such a complete analysis has been investigated for GR (cf. Damour~\cite{Damour87} and references therein), but it is very complex. In my opinion, it would be more convincing to find a straightforward justification of textbook calculations~\cite{Will,Weinberg,L&L,MTW} that derive the quadrupole formula and that use it directly to assess the Newtonian energy loss in the emitting system.\\

This is what I have tried to do for the scalar theory, by proposing an ``asymptotic post-Minkowskian (PM) scheme"~\cite{A34,B24}, which is the equivalent, for problems of gravitational radiation, of the asymptotic PN scheme summarized in Subsection \ref{PNA}, the latter being adapted to usual problems of celestial mechanics. Thus, that PM scheme also consists in introducing a family of initial conditions, which define a family of perfect-fluid gravitating systems. The essential difference between the PN and the PM scheme is that, in the PM scheme, the initial condition for the velocity field does not depend on the field-strength parameter $\varepsilon$, which means physically that it is not {\it necessary} for the velocities of the bodies to be negligible with respect to $c$. (The PM scheme is thus simply {\it more general} than the PN scheme.) Due to this fact, the expansions with respect to $\varepsilon$ are valid at a fixed value of the ``real" time (that mesured in units independent of $\varepsilon$). In turn, this implies that the relevant time variable for the expansions in powers of $1/c^2$ (which assume the same $\varepsilon$-dependent change of units as in the PN case) is not any more $x'^0\equiv T$, but instead $x^0\equiv cT$. It follows that the equation (\ref{champ2}) for the scalar field is expanded to wave equations involving d'Alembert's operator, instead of the Laplace operator that appears in the Newtonian limit. Neither does the expansion of the dynamical equations lead to the Newtonian equations: instead, the 0-order equations express the {\it local conservation} of momentum (and mass-energy). In the scalar theory, that disposes of a true conservation law for energy, one may unambiguously calculate the rate of the total (material plus gravitational) energy which is contained in an isolated domain. This rate is expressed by an outgoing gravitational flux. The 0PM expansion of that flux depends on the retarded potential, solution of the relevant wave equation obtained by expanding (\ref{champ2}). As it is also the case in GR under the harmonic gauge, the limiting value of the flux depends only on the third time derivative of the quadrupole tensor. This turns out to be true even if the mass center of the system is not at rest in the preferred reference frame~\cite{A34}.\\

This result can be used to approximate the rate of the Newtonian energy of the isolated weakly-gravitating system, because this PM scheme is more general than the PN scheme~\cite{B24}. One may then calculate the so-called ``Peters-Mathews coefficients" of the scalar theory. These do not have the same values as in harmonic GR, but they give an energy loss, hence a decrease in the orbital period, having the same order of magnitude as in GR---if one takes values for the orbital parameters and the masses that are similar to the values found~\cite{TaylorWeisberg} by adjusting on binary pulsar data the ``timing model" based on GR~\cite{A34,B24}. Now the observational data of a such model is merely the list of the reception times of the pulses emitted by the pulsar. Therefore, one cannot say in advance what would be the exact values of the orbital and mass parameters that would be found by adjusting the pulse data using a timing model entirely based on the scalar theory. Thus, at the current stage, one may state that {\it the scalar theory does predict a decrease in the orbital period of a binary pulsar, with the same order of magnitude as in harmonic GR---contrary to virtually all alternative theories, except for those which are an extension of GR}~\cite{Will}. It would remain to investigate whether the pulse data may be fitted with the same accuracy as they can be in harmonic GR.

\section{Link with classical electromagnetism and with quantum theory}

\subsection{Maxwell equations in a gravitational field} \label{Maxwell}

In GR, the transition from the usual Maxwell equations, which are valid in a flat space-time, to those that apply in a gravitational field, thus in a curved space-time, is obtained by applying a simple rule: substitute the covariant derivatives with respect to the curved metric $\Mat{\gamma}$, in general space-time coordinates, for the partial derivatives in Galilean coordinates (which are covariant derivatives with respect to the flat metric, in those Galilean coordinates). Note, however, that this rule fixes the equations only if one starts from the equations for the electromagnetic field tensor, not from those for the 4-potential; indeed the latter ones are second-order, but covariant derivatives do not commute. This difficulty is serious for wave equations such as the Klein-Gordon equation, because there is not always a ``natural" first-order equation to be associated with. The assumption of ``minimal coupling," according to which the curvature should not play a role ``unless there is a good physical reason," looks like an expedient, ad hoc prescription. Anyway, the investigated theory is not a ``metric theory"~\cite{Will}, because free test particles generally do not follow geodesic lines of metric $\Mat{\gamma}$ (although the latter is related to physical space and time measurements in the same way as in GR). One has, therefore, to find another way to extend Maxwell's equations to the situation with gravitation in this theory.\\

In this theory, the first group of the Maxwell equations is that obtained in the standard way from the definition of the electromagnetic field $\mathbf{F}$ by a 4-potential, while the second group results from the dynamics of a charged medium subjected to gravitation and to the Lorentz force due to the electromagnetic field (Sect. 7 in Ref.~\cite{B13}). The extension to the scalar theory of Newton's second law for a test particle (Subsection \ref{ExtensNewton}) can be formulated also in the case that the particle is subjected to a nongravitational force, in addition to the gravitational force. As in the ``purely gravitational" case, one then deduces an energy equation for the test particle, and afterwards one obtains the expression of its 4-acceleration. Considering then a dust of such particles, a dynamical equation for a continuum is got in the same way as in the purely gravitational case (Subsection \ref{C.dynamics}). The r.h.s. of that equation is the sum of $b_\mu$ (Eq.~(\ref{continuum})) and of a term that involves the density of the nongravitational force, assumed given. It is again admitted that this equation holds true for a general continuum, thus extending continuum dynamics to the case with a density of external nongravitational force.\\

In order to apply this dynamics to a charged medium, one determines first the expression of the Lorentz force by demanding that it be invariant under the coordinate changes which are internal to the preferred frame, and that it reduce to the standard expression of SR if the gravitational field cancels. Thus, the charged medium, with energy-momentum tensor $\mathbf{T}_\mathrm{charges}$, obeys the continuum dynamics for a medium subjected to the density of the Lorentz force. Now the total energy-momentum tensor is the sum $\mathbf{T}=\mathbf{T}_\mathrm{charges}+\mathbf{T}_\mathrm{em}$, where $\mathbf{T}_\mathrm{em}$ is the standard energy-momentum tensor associated with the e.m. field $\mathbf{F}$. The total tensor $\mathbf{T}$ must obey the dynamical equation in the absence of non-gravitational forces, Eq.~(\ref{continuum}). But since the r.h.s. $b_\mu$ of~(\ref{continuum}) is linear in $\mathbf{T}$, these two dynamical equations (that for $\mathbf{T}_\mathrm{charges}$ and that for $\mathbf{T}$) imply a dynamical equation for tensor $\mathbf{T}_\mathrm{em}$, which may be interpreted in saying that {\it the e.m. field is a continuous medium subjected to the gravitation and to the opposite of the Lorentz force}. By virtue of the known expression of $\mathbf{T}_\mathrm{em}$ as function of the e.m. field $\mathbf{F}$ and the metric $\Mat{\gamma}$, the dynamical equation for tensor $\mathbf{T}_\mathrm{em}$ provides the four additional equations sought for~\cite{B13}. They are nonlinear, but they reduce to the usual Maxwell equations of metric theories, in the particular case where the gravitational field is constant.

\subsection{Identity of the motions of light waves and light particles}

We must find the necessary link between the trajectories of ``photons" and the modified Maxwell equations, thus between geometrical and wave optics in a gravitational field. In other words, the theory has to say which particular kind of electromagnetic field can be regarded as a ``dust of photons," each of which follows, by definition, a trajectory at the constant velocity $c$, solution of Newton's second law for a particle with no rest-mass, subjected to a purely gravitational force. The modified Maxwell equations outlined above express the dynamics of the e.m. field subjected to gravitation and to the opposite of the Lorentz force. However, in addition to these two external forces, the e.m. field continuum will generally be subjected also to internal forces, like any other continuum. In order for its dynamics to reduce locally to that of a free photon, it is necessary that both the Lorentz force and those internal forces cancel. The cancellation of the Lorentz force is equivalent to that of the 4-current, thus to considering a region outside the charged medium, that is, ``in vacuo," in the sense of classical physics.\\

It is less immediate to identify the internal forces and to find the conditions under which they cancel. The following method allows one to identify the internal forces in a general continuum: first, one writes the dynamics of this continuum in the form of Newton's second law for a volume element, which is thus subjected to both external and internal forces. Second, one demands that this dynamics coincides with the equation obtained, as outlined in Subsection \ref{Maxwell}, by induction from a dust. One then finds that, in order that the internal forces cancel, it is necessary and sufficient that the $\mathbf{T}$ tensor be the tensor product of a 4-vector $\mathbf{V}$ by itself, 
\be    \label{tensorsquare}
\mathbf{T} = \mathbf{V}\otimes \mathbf{V}.
\ee
This is the general form of the energy-momentum tensor of a ``dust," independently of any assumption about the constitutive particles and their rest-mass. Moreover, under condition (\ref{tensorsquare}), the absolute velocity of the continuous medium is unambiguously defined as the velocity of the energy flux:
\be   \label{energyflux}
u^i \equiv \frac{\dd x^i}{\dd T} = c \frac{T^i_0}{T^0_0}.
\ee 
For the energy-momentum tensor $\mathbf{T}_\mathrm{em}$, of the e.m. field, assuming condition (\ref{tensorsquare}) is equivalent to imposing that the two invariants of the e.m. field are zero~\cite{Stephani82}, which characterizes a ``pure electromagnetic wave".\\

Thus, in order that the e.m. field behave as a ``dust of photons," it is necessary and sufficient that its two invariants cancel and that one considers a region free of charges. In that case, one shows~\cite{B13} that the trajectories of the e.m. energy flux, defined by Eq.~(\ref{energyflux}) with $\mathbf{T}=\mathbf{T}_\mathrm{em}$, are photon trajectories in the sense of Newton's second law of the scalar theory. This provides the necessary link which was sought for.

\subsection{Interpretation of the Hamiltonian-wave equation correspondence} \label{correspondence}

If ones seriously aims at making quantum theory and the theory of gravitation match together, one should not blindly accept the formalism of quantum theory (not more than that of GR!) and try to formally adapt it to the situation with gravitation. Instead, one should make attempts at understanding the foundations of quantum mechanics (QM). Perhaps the most important and mysterious feature of quantum theory (including quantum field theory) is the ``quantum correspondence," that associates a linear differential operator with a classical Hamiltonian, and that leads to regard energy and momentum as operators of this kind. It seems that there is no real explanation of that correspondence to be found in the literature. However, it turns out~\cite{Whitham} that, for a linear wave equation, the wave vector propagates along the bicharacteristics of a certain linear partial differential equation (PDE) of the first order. When the latter equation is put in characteristic form, one obtains a Hamiltonian system, in which the Hamiltonian is none other than the dispersion relation associated with the wave equation. Now, for a given ``wave mode," the correspondence between the dispersion relation and the wave equation is one-to-one. These remarks were made by Whitham~\cite{Whitham} in the context of the theory of classical waves and they provide the basis for my attempt~\cite{B15,A22} at interpreting the quantum correspondence.\\

To begin with, it is worth to generalize somewhat Whitham's remarks. The notion of a wave assumes that the wave function $\psi$ is defined over an open domain D in an extended configuration space $\mathrm{V}\equiv\mathbf{R}\times \mathrm{M}$, with $\mathrm{M}$ the $N$-dimensional configuration space. Consider a linear differential operator P defined ``on D" (in fact, defined on a suitable space of functions, each of which is defined on D). For any $X\in\mathrm{D}$, a polynomial function $\Pi_X$ of covector $\mathbf{K}\in\mathrm{T^*V}_X$ can be naturally defined with the coefficients of P. 
\footnote{
~The coefficients of P are coordinate-dependent. It turns out that $\Pi_X$ is well-defined if and only if one can define a certain class of coordinate systems connected by ``infinitesimally linear" changes, that is,
\be \label{infinit-linear}
\frac{\partial ^2x'^\rho}{\partial x^\mu\partial x^\nu}=0
\ee
at the point $X((x_0^\mu))=X((x_0'^\rho))$ considered, and if one admits only those coordinate systems that belong to this class. In particular, if the space $\mathrm{V}$ is endowed with a pseudo-Riemannian metric $\Mat{\gamma}$, then $\Pi_X$ stays invariant if one considers only coordinate systems that are locally geodesic at $X$, i.e., $\gamma_{\mu \nu,\rho}(X)=0$ for all $\mu,\nu,\rho$---the change from any such system to another one verifies indeed (\ref{infinit-linear}).
}
The correspondence between the linear operator P and the function $\Pi$ is {\it one-to-one}. The inverse correspondence consists in replacing the component $K_\mu$ of $\mathbf{K}$ by the operator $D_\mu/i$, where $D_\mu$ is the partial derivative with respect to $x^\mu$ and $i=\sqrt{-1}$. Suppose one is able to follow as function of $X$ the different roots of the dispersion equation $\Pi_X(\mathbf{K})=0$, seen as a polynomial equation for the frequency $\omega\equiv-K_0$, thus one is able to identify the different wave modes. Then it is possible to define different ``dispersion relations" 
\be \label{dispersion}
\omega = W(K_1,...,K_N;X),
\ee
each of which gives the frequency $-K_0$ as a function of the ``spatial wave (co)vector" $\mathbf{k}\equiv (K_1,...,K_N)$, i.e. the spatial part of $\mathbf{K}$, for the considered wave mode. For each of these functions $W$, Whitham's reasoning shows that a wave function of the form $\psi (X) = A(X)\, \mathrm{exp}[\, i\,\theta (X)]$, whose the ``wave covector" $\mathbf{K}\equiv \mathrm{grad}\,\theta $ is a solution of Eq.~(\ref{dispersion}), allows one to define naturally a field of trajectories which are solutions of the Hamiltonian system with Hamiltonian $W$.\\

When they invented wave mechanics, de Broglie and then Schr\"odinger considered that a Hamiltonian system describes the ``skeleton" of a wave pattern associated with a linear wave equation, in the same way as geometrical optics describes the trajectories of light rays, which are the skeleton of the underlying wave pattern. If one adopts this heuristics, then the mathematical framework which has been summarized hereabove leads really to admit that the wave equation, which is searched for, must have as one of its dispersion relations a function $W$ whose Hamiltonian trajectories are just the solution trajectories of the classical Hamiltonian $H$, from which one starts. In order to ensure this, the most natural way is to assume that $W$ and $H$ are proportional. Denoting $\hbar$ the proportionality constant, one gets simultaneously the ``quantum" relations that relate energy with frequency and momentum with wave vector, and the correspondence between a classical Hamiltonian and a wave operator~\cite{B15,A22}. This is quite striking. Moreover, this interpretation provides a resolution to the ambiguity which is inherent in the correspondence, in the general case where the Hamiltonian contains terms that depend on both the ``position" $X$ (in the extended configuration space) and the canonical momentum. It also indicates a few tracks to reflect on quantum mechanics, especially on its extension in a gravitational field (Subsection \ref{gravitKG} below).

\subsection{Klein-Gordon wave equation in a gravitational field} \label{gravitKG}

Even before one tries to quantize gravitation, one should first solve the important and observationally relevant problem of extending quantum mechanics of nongravitational particles and fields to the situation with a gravitational field. This needs that a framework should be found, in which one could extend the quantum wave equations (and what is done with them, e.g. finding stationary energy levels, etc.) to a curved space-time. One then has to ask which is the relevant covariance group for these wave equations. The suggested interpretation of the quantum correspondence in the framework of Schr\"odinger's wave mechanics (Subsection \ref{correspondence}) leads to strongly restrict the choice in the coordinate system~\cite{B15}. Indeed, in order that function $\Pi$ be unambiguously defined, it is necessary to limit oneself to coordinate changes that are ``infinitesimally linear," Eq.~(\ref{infinit-linear}). Moreover, the role played by Hamiltonian systems, as also the very definition of the group velocity, which is important in this interpretation~\cite{B15,A22}, impose that one makes a clear distinction between space and time, the time coordinate being in fact fixed up to a mere scale change. In a flat space-time, this leads to use Galilean coordinates in an a priori chosen inertial frame, and it fortunately happens that one thus gets Lorentz-invariant equations, provided one starts from a Lorentz-invariant Hamiltonian. In a theory of gravitation in a curved space-time, however, it is in general impossible to restrict the choice of the coordinate system so as {\bf i}) to fix the time coordinate and {\bf ii}) to pass from one system to another one by an infinitesimally linear coordinate change---unless one imposes this restriction in an arbitrary way, which is physically meaningless. This is only possible in a preferred-frame theory, such as the theory investigated.
\footnote{
~The suggestion to consider a fixed reference frame (at least in a first step) appears even independently of any interpretation of the quantum correspondence, as one may show~\cite{B17} by studying the simplest relativistic wave equation, i.e., the Klein-Gordon equation. If one wishes to extend that equation to a curved space-time, one may first try to apply the standard quantum correspondence to the expression of the invariant length of the 4-momentum. But the wave equation obtained thus will depend on the coordinate system. Even in a given system, there will be an ambiguity arising from the fact that position and momentum operators do not commute. Second, one may think of merely transposing the writing of the wave equation, substituting covariant derivatives for partial derivatives---but this also is ambiguous, due to the fact that covariant derivatives do not commute.
} 
In a such theory, one class of coordinates verifying conditions {\bf i}) and {\bf ii}) appears naturally: this is the class of the coordinate systems which are bound to the preferred frame, locally geodesic for the spatial metric in that frame, and whose time coordinate is, up to a constant factor, the preferred time coordinate (the ``absolute time" $T$, for the scalar theory)~\cite{B15}. However, in the particular case of a {\it static} metric, one has for any metric theory a preferred frame and a preferred time coordinate---namely, the frame in which the metric is static, which also fixes the time coordinate.\\

Thus, when applied to to the situation with gravitation, our interpretation of wave mechanics suggests a restriction on the admissible coordinate systems, which leads one either to consider a preferred-frame theory, or to study only the static case. It was then natural to ask whether or not the motion of a free test particle in the scalar theory derives from a classical Hamiltonian. The answer~\cite{B15} is that the energy $e$ of the particle with rest-mass $m$ in the gravitational field $\beta$,
\be \label{energy}
e \equiv \beta\gamma_v m c^2, \quad \beta = \sqrt{\gamma_{00}}
\ee
\cite{A15}, is indeed a Hamiltonian for its dynamics as defined by the extension of Newton's second law, if and only if the gravitational field is constant---hence in fact static, in view of Eq.~(\ref{gamma})$_3$. ($\gamma_v$ is the Lorentz factor, $v$ being of course evaluated with the physical metric.) But, in the static case, that dynamics is identical to geodesic motion in the static space-time metric $\Mat{\gamma }$, independently of any particular form of that static metric~\cite{A16}. Therefore, the Hamiltonian (\ref{energy}) applies to any metric theory in the static situation. Thus, for a massive test particle in the static case, in the scalar theory as also in any metric theory, one may apply the unambiguous quantum correspondence suggested by the proposed interpretation of wave mechanics. One obtains~\cite{B15} the following extension of the free Klein-Gordon equation to a static gravitational field:
\be \label{gravit-KGeq}
\Delta_\Mat{g} \psi - \frac{1}{c^2} \frac{\partial ^2 \psi}{\partial t_{\mathbf{x}}^2}= \frac{1}{\lambda^2}\psi , \quad \lambda \equiv \frac{\hbar}{mc}, \quad \frac{\partial }{\partial t_{\mathbf{x}}} \equiv \frac{1}{\beta (t,\mathbf{x})}\frac{\partial }{\partial t},
\ee
where $t$ is the preferred time of the static metric. It is striking that the wave operator on the left-hand side is none other than that which appears in the equation (\ref{champ1}) for the gravitational field in the scalar theory! In the framework of that theory, it seems natural to assume that Eq.~(\ref{gravit-KGeq}) is valid in the general case, but then it is of course a preferred-frame equation, which applies in the ether frame E.\\

In summary, the analysis of the mathematical meaning of Schr\"odinger's wave mechanics leads us to find that the presence of a gravitational field might break Lorentz invariance by making a preferred reference frame appear, in just the same way as it is the case in the investigated scalar theory. I also mention that, according to the concept of gravity as due to the macroscopic pressure gradient in the constitutive ether (Subsect. \ref{grav=pression}), {\it gravitation does not have to be quantized,} see Sect. 5 in Ref.~\cite{B15}. The problem is then ``merely" to write quantum mechanics (and quantum field theory) in a gravitational field. It is hoped that the foregoing is a sound way to attack this task.

\section{Conclusion}

The main motivation for developing an ether theory of gravity, in the line of the Lorentz-Poincar\'e version of special relativity, has a philosophical nature: it is the belief that space and time are basically separated entities, and that objects cannot interact through empty space---even less, of course, if that interaction has to be instantaneous. We, the defenders of this point of view, are not necessarily unable to play with abstract mathematical concepts. To take a simple example, it is admitted by us that the concept of the space-time is a very clever and convenient invention, which fits very well to handle the mathematics of relativistic theories. Starting from Lorentz-Poincar\'e SR with its homogeneous and isotropic ether, and trying to build an understandable explanation for gravity, one is unavoidably led to think that this phenomenon must be the manifestation of some heterogeneity of the ether. Thus, Podlaha \& Sj\"odin interpret the post-Newtonian effects as being basically due to a spatial variation in the ``ether density." (A few of these effects were predicted on the basis of Einstein's general relativity {\it and then} observed, although two extremely important ones: the advance in perihelion and the light deflection, were respectively observed {\it or} (half-)predicted before that.) The ``ether density" is seen as a refractive index, and it plays a role similar to that played by the square of the velocity in SR. Winterberg emphasizes the role of the absolute velocity itself, i.e., he considers that, just like in Lorentz-Poincar\'e SR, the post-Newtonian gravitational effects are due to our motion through the ether, his ether having a complex fluid motion with respect to an inertial frame in the gravitational case. Dmitriyev assumes that gravitation is a propagation of dilatational point defects in a crystal-elastic substratum, thus involves density fluctuations, as for Consoli.\\

My own attempt started from Romani's concept of a fluid ``constitutive ether," in which gravity was interpreted as the gradient of ether density. At the heuristic level it is hence closest to the Podlaha-Sj\"odin concept, though it differs from the latter (and from Romani's concept) already at this level---in particular, gravity is seen as a pressure force (Archimedes' thrust) and it can be characterized by a {\it decrease} in the ether density, not an increase, towards the attracting bodies. But the main difference is that I could develop my attempt farther. It is a complete theory of gravitation, including the necessary links with electromagnetism and with quantum theory. The latter link is at the stage of a beginning, but it seems to constitute a firm basis. It appears from several reasons that the presence of gravity might make a preferred reference frame appear in the equations of quantum mechanics. This possibility is clearly incompatible with GR. But it is fully compatible with the investigated theory---to the point that the same wave operator appears, for a priori independent reasons, in the extension of the Klein-Gordon equation and in the equation for the scalar gravitational field.\\

As to the experimental comparison: the gravitational effects on light rays are the standard effects predicted by GR. The energy loss by gravitational radiation has a form very similar to the one it has in GR under the harmonic gauge, hence it is plausible that the pulse data emitted by binary pulsars may be also accurately fitted. An important work has been done to develop a consistent approximation scheme for celestial mechanics, in the line of the asymptotic schemes used in applied mathematics. A similar work for GR exists only in an incomplete form, and it could be difficult to develop it up to a stage equivalent to that which has been reached here, because the constraint equations of GR do not allow to define easily a family of initial data from that corresponding to a general-enough gravitating system. In GR, the agreement with astrodynamical observations is based on using EIH-like equations, but it is not clear the extent to which the predictions obtained with these equations agree with what the exact theory would predict. In particular, I do not see arguments showing convincingly that the internal structure of the bodies should not play a role in GR, as it does in the scalar theory. In the latter, the structure influence subsists at the point-particle limit. There are indications that this violation of the weak equivalence principle (WEP) might also happen in GR in ``Schwarzschild-like" gauges, but it would be more difficult to check it. The celestial mechanics of the scalar theory has been adjusted on an {\it ephemeris} based on standard post-Newtonian equations of GR (EIH-like equations). The residual difference with the ephemeris is not negligible, but I argue that one can hope that an adjustment on {\it primary observations} would allow a sensible matching of these. Finally, the scalar theory predicts that the cosmic expansion {\it must} be accelerated, instead of just accommodating the observed acceleration by introducing new parameters. It also predicts an interesting new phenomenon: matter creation/destruction in a variable gravitational field.\\

{\bf Post-Scriptum.} Apart from relatively minor changes, this work dates back to January 2004 (arXiv:gr-qc/0401021v1). Since then, a modified version of the scalar ether-theory has been built \cite{O3,GAS04,A35}, based on an isotropic space metric. This eliminates \cite{A35} the WEP violation found with the present version, which is based on an anisotropic space metric. Moreover, the asymptotic scheme of Futamase \& Schutz for GR in the harmonic gauge \cite{FutaSchutz} has been studied and developed up to the obtainment of tractable equations of motion \cite{A36}, see Note 17 here, in \S \ref{!!!}. Finally, the method summarized in Subsect. \ref{gravitKG} has been adapted to the Dirac equation: the Dirac equation has been derived directly from wave mechanics, i.e., from the classical Hamiltonian \cite{A37}. This derivation applies also to a gravitational field, be it a static one \cite{A37} or a general one \cite{A39}. However, {\it two} distinct gravitational Dirac equations may be thus derived, none of which does coincide with the standard gravitational Dirac equation, that is due to Weyl and to Fock \cite{A39}.

\bigskip

{\bf Acknowledgements.} I would like to thank Dr M. C. Duffy, who initiated and organized the ``Physical Interpretations of Relativity Theory" conference series. This series has provided a strongly-needed forum for researchers who do not give up searching for a clearer physics. I am still grateful to Dr Duffy for his initiative of the forthcoming book: ``Ether, Spacetime and Cosmology," which will contain the present contribution. Many thanks to Franco Selleri, who encouraged this research, most recently in harboring me in the Physics Dept. of the Bari University for a one-year stay; thanks also to Professor F. Romano for that matter. Last but not least, I would like to express my gratitude to my colleagues of the ``Solid Mechanics" section of the CNRS, as well as to the successive directors of my laboratory and to other colleagues in that lab, for their tolerance to this research---which is quite remote from the other themes in that field.

\bibliographystyle{amsplain}

\end{document}